\documentclass[10pt,journal,compsoc]{IEEEtran}
\ifCLASSOPTIONcompsoc
  \usepackage[nocompress]{cite}
\else
  \usepackage{cite}
\fi
\ifCLASSINFOpdf
\else
\fi
\usepackage{enumitem}
\usepackage{amsmath}
\usepackage{amsfonts}
\usepackage{multirow}
\usepackage{tabu}
\usepackage{makecell}
\usepackage{wrapfig}
\usepackage{color}
\usepackage{booktabs}
\usepackage{ifthen}
\usepackage{hyperref}
\usepackage{subcaption}
\usepackage{bbold}
\usepackage{soul}
\usepackage[linesnumbered,ruled,vlined]{algorithm2e}
\usepackage{mathtools}
\usepackage{gensymb}

\usepackage{xcolor}
\usepackage[normalem]{ulem} 
\definecolor{gray}{rgb}{0.5,0.5,0.5}
\definecolor{green}{rgb}{0, 0.6, 0}
\definecolor{orange}{rgb}{1, 0.5, 0}
\definecolor{mahogany}{rgb}{0.75, 0.25, 0.0}
\definecolor{purple}{rgb}{0.6, 0, 0.6}
\definecolor{darkgreen}{rgb}{0, 0.3, 0}
\definecolor{orange}{rgb}{1, 0.5, 0.}

\newcommand{\majorhl}[1]{#1}

\newcommand{\ignore}[1]{}
\newcommand{\none}[1]{}
\newcommand{\com}[1]{}

\newcommand{\shortcite}[1]{\cite{#1}}

\newcommand{\etal}{{\it{et~al.}}}
\newcommand{\ie}{i.e.,}
\newcommand{\eg}{e.g.,}
\newcommand{\figname}{Figure}
\newcommand{\tabname}{Table}
\newcommand{\secname}{Section}

\newcommand{\eqname}{Eq.}

\begin{document}
%
\title{ClipGen: A Deep Generative Model for Clipart Vectorization and Synthesis}
%
%
%
%

\author{I-Chao Shen~\IEEEmembership{}and~Bing-Yu Chen,~\IEEEmembership{Senior~Fellow,~IEEE}
\IEEEcompsocitemizethanks{
\IEEEcompsocthanksitem I-C. Shen, and B-Y. Chen are with with National Taiwan University.\protect\\
E-mail: jdilyshen@gmail.com, robin@ntu.edu.tw
}
\thanks{Manuscript received April 19, 2005; revised August 26, 2015.}}
\markboth{Journal of \LaTeX\ Class Files,~Vol.~14, No.~8, August~2015}%
{Shell \MakeLowercase{\textit{et al.}}: Bare Demo of IEEEtran.cls for Computer Society Journals}
\IEEEtitleabstractindextext{%
\begin{abstract}
This paper presents a novel deep learning-based approach for automatically vectorizing and synthesizing the clipart of man-made objects.
Given a raster clipart image and its corresponding object category (\eg~airplanes), the proposed method sequentially generates new layers, each of which is composed of a new closed path filled with a single color.
The final result is obtained by compositing all layers together into a vector clipart image that falls into the target category.
The proposed approach is based on an iterative generative model that (i) decides whether to continue synthesizing a new layer and (ii) determines the geometry and appearance of the new layer.
We formulated a joint loss function for training our generative model, including the shape similarity, symmetry, and local curve smoothness losses, as well as vector graphics rendering accuracy loss for synthesizing clipart recognizable by humans.
We also introduced a collection of man-made object clipart, \textit{ClipNet}, which is composed of closed-path layers, and two designed preprocessing tasks to clean up and enrich the original raw clipart.
To validate the proposed approach, we conducted several experiments and demonstrated its ability to vectorize and synthesize various clipart categories.
We envision that our generative model can facilitate efficient and intuitive clipart designs for novice users and graphic designers.
\end{abstract}
}

\maketitle

\IEEEdisplaynontitleabstractindextext

%
\IEEEpeerreviewmaketitle

\section{Introduction}
\label{sec:intro}
Vector clipart is widely used in graphic design to express concepts or illustrate daily life objects compactly. It has several advantages over raster images, such as (i) better geometric editability benefit from separate geometric paths and (ii) resolution independence, which significantly reduces the burden of redesigning the same concept or shape whenever a new resolution support is required.
However, designing vector clipart from scratch or editing existing clipart obtained from online repositories ( \eg~\textit{Openclipart}\footnote{http://clipart-library.com/openclipart.html}) is challenging for amateur users who lack the skill of using vector graphics editing software, such as \textit{Adobe Illustrator}~\cite{illustrator} and \textit{Inkscape}~\cite{Inkscape}.
To address this problem, an automated method to synthesize clipart for amateur users is required.
Deep generative image modeling techniques, which enable the automatic generation of high-quality raster images for human faces, animals, and natural objects, are being rapidly developed~\cite{karras2019analyzing}.
However, their synthesized results are raster images that lack the structural editability feature available in vector clipart.

Therefore, this paper focuses on generating vector clipart to support higher-level structural editing operations.
Our goal was to train a deep generative model that can simulate the design process of vector clipart.
In particular, we focused on designing a generative model to represent and generate clipart conditioned on a target raster image and the desired object category sequentially.
The proposed method is based on two essential clipart characteristics.
First, the vector clipart is defined to comprise separate layers containing a single closed path and its filled color.
Second, the clipart geometries exhibit strong symmetry and simplicity for better editability.
On the basis of these two characteristics, we propose an iterative clipart synthesis framework that synthesizes a layer with a predicted single closed path and its filled color in one step.
Finally, we composite all the synthesized layers to obtain the final result (please refer to the examples presented in Figs.~\ref{fig:first_syn} and \ref{fig:second_syn}).

To describe the composited result at the current step, we extract several visual representations, including the types of existing curves, their depth ordering, the location of their control points, and the rendered image.
We then use convolutional neural networks (CNNs) to recognize and extract the essential features of these visual representations.
The extracted features capture patterns and relationships between curves, which provide a useful context for deciding which curve to synthesize next.
On the basis of previous works on 3D shapes~\cite{kalogerakis2012probabilistic} and 3D scene synthesis~\cite{wang2018deep}, we model an iterative synthesis process using two separate modules. 
Given the visual representations of the current synthesized result, the first module (i) decides whether to synthesize a new layer and (ii) determines the centroid location probability of the next closed path, given the visual representations of the current synthesized result.
The second module is a critical component of our iterative synthesis framework because it decides the geometry and the appearance of the next layer.
We use recurrent neural networks (RNNs) to predict the sequence of control points of the connected curves that forms a closed path.
In addition, we design several loss functions, including shape loss, symmetry loss, area loss, local smoothness loss, and simplicity losses, to aid the network in synthesizing the desired curves.
We also adopt a rendering accuracy loss based on a novel differentiable vector graphics rendering technique~\cite{Li:2020:DVG}, which encourages the network to refine the predicted geometry and predict the appropriate filling color simultaneously.

To facilitate our clipart generative model training, we introduce a clipart dataset called \textit{ClipNet}, which contains \majorhl{2000} clipart classified into ten categories of man-made objects.
This dataset focuses on vector graphics representation and thus complements the previous image~\cite{imagenet_cvpr09} and 3D shape~\cite{chang2015shapenet} datasets. 
The characteristics of \textit{ClipNet} are analyzed, and two annotation tasks are designed to extract clean shapes from the raw clipart collected from online repositories.
\textit{ClipNet} can be used in several potential applications, such as clipart generative and recognition models as well as multi-view clipart synthesis.

We applied the proposed method to several applications, such as vectorizing existing raster clipart, synthesizing various categories of novel clipart, and synthesizing clipart that resembles a photograph.
\majorhl{
These applications can facilitate the clipart design process that involves the use of common categories of man-made objects.
}
We conducted ablation studies to validate our method and compared the results of our method with those of existing methods.

\section{Related Work}
\label{sec:related}

\subsection{Image vectorization and clipart synthesis}
The vectorization of raster images is a long-standing research problem.
Commercial products~\cite{illustrator, vector_magic} enable robust natural image vectorization that can simultaneously address both image segmentation and curve (segment boundary) fitting problems.
Many studies have been conducted on image segmentation part~\cite{FLB17,orzan2008diffusion,sun2007image,xia2009patch,kim2018semantic} and curve fitting part~\cite{bessmeltsev2019vectorization}.
Favreau~\etal~\cite{FLB17} focused on vectorizing photographs in cliparts with easily editable geometries and representing the raster image accurately.
Kim~\etal~\cite{kim2018semantic} used a neural network to predict a probability map of pixel-path relationships and formulated a global MRF problem to segment an input image into different paths.
Liu~\etal~\cite{Liu:2016:DI} designed an interactive system to synthesize novel clipart by remixing the clipart existing in a large repository.
By contrast, Xie~\etal~\cite{xie2017} provided interactive approaches that allow users to adjust region boundary properties manually.
Unlike the approaches used in the aforementioned studies, the approach proposed in the current study can synthesize man-made object clipart without conditioning on any specific input.
\majorhl{Moreover, the proposed approach can synthesize filled images instead of line drawings without filled colors.
}

Some studies have focused on vectorizing pixel arts~\cite{hqx,kopf2011depixelizing,hoshyari2018perception}.
For example, Kopf~and~Lischinski~\shortcite{kopf2011depixelizing} proposed a dedicated method focused on resolving the topological ambiguities that occur during vectorizing pixel arts.
Hoshyari~\etal~\shortcite{hoshyari2018perception} has used human perceptual cues to generate a better boundary vectorization that better matches viewers' expectations.

\subsection{Generative model}
In computer graphics, the generative models for 3D models and scenes have drawn considerable research attention.
Many assembly-based 3D modeling methods provide a probabilistic model to infer potential components for assembling under fully-automatic~\cite{kalogerakis2012probabilistic} and semi-automatic~\cite{chaudhuri2011probabilistic} settings.
Similar probabilistic modeling methods have been used to synthesize novel 3D scenes~\cite{fisher2012example} and suggest pattern colorizations~\cite{lin2013probabilistic}.

Deep generative models have gained popularity for synthesizing digital content, including raster images, videos, and audio.
The most popular deep generative models include variational autoencoder (VAE)~\cite{kingma2013auto}, which aims at maximizing the lower bound of the data log-likelihood, and generative adversarial network (GAN)~\cite{goodfellow2014generative}, which aims to achieve an equilibrium between the generator and the discriminator.
Many variations of the aforementioned models have been used for the unconditional and conditional generation of images, including for image translation~\cite{pix2pix2016,CycleGAN2017}.

The progresses achieved in the raster image domain has inspired researchers to synthesize 3D models~\cite{li2017grass,zhu2019scores}, 3D scenes~\cite{wang2018deep,li2019grains}, 3D abstractions~\cite{zou20173d}, and 3D motions for man-made objects~\cite{Yan:2019:RRP}.

Moreover, the deep generative models have facilitated tasks such as 2D sketches~\cite{ha2017neural}, 2D design layouts~\cite{zheng2019content,li2018layoutgan}, interactive annotations on 2D images~\cite{acuna2018efficient}, pixelization~\cite{han-2018-pixelization}, font exploration~\cite{Lopes_2019_ICCV}, and face image decomposition~\cite{sbai2018vector}.
The learned latent representations of these models are useful for creative tasks such as garment design~\cite{garmentdesign_Wang_SA18} and terrain design~\cite{guerin2017interactive}.
The latent space can be explored through predefined interactions~\cite{Bau:Ganpaint:2019} or free-form interactions~\cite{hin2019interactive,zhu2016generative} for synthesizing images that meet the users' requirements.

\subsection{Image and shape dataset}
Advances have been achieved in the recognition and generative applications of deep neural networks by using the image, video, 3D shape, and audio datasets.
In the raster image domain, ImageNet~\cite{imagenet_cvpr09} formed the foundation for this wave of neural network revival.
Later, many task-specific datasets, such as COCO~\cite{lin2014microsoft}, CelebA~\cite{liu2015deep}, and DAVIS~\cite{Perazzi2016}, were published to support tasks such as image or video segmentation and captioning, object detection, and face synthesis.
Moreover, many 3D models and scenes datasets, such as ShapeNet~\cite{chang2015shapenet}, ModelNet~\cite{Wu_2015_CVPR}, PartNet~\cite{Mo_2019_CVPR}, ABC~\cite{Koch_2019_CVPR}, and SUNCG~\cite{song2016ssc}, have made the 3D deep learning in recognition and generative applications more feasible.
Datasets in the creative domain, such as Creative Flow+~\cite{shugrina2019creative}, provide 
diverse multi-style artistic video renderings with labeled per-pixel metadata.
OpenSketch~\cite{GSHPDB19}, QuickDraw~\cite{ha2017neural}, TUBerlin~\cite{eitz2012hdhso}, and other databases~\cite{Cole:2008:PDL} provide sketches created by professional artists with different taxonomies of line types and viewpoints.

Inspired by ShapeNet~\cite{chang2015shapenet}, in this paper, we introduce \textit{ClipNet}, which is composed of clipart of different categories of man-made objects, such as chairs, airplanes, and cameras.
Clipart is created using vector graphics editors, such as Adobe Illustrator~\shortcite{illustrator} or Inkscape~\shortcite{Inkscape}, and integrates both geometry and appearance into a compact format with simplified editability.
These two characteristics distinguish this clipart data from both raster images and 3D models.
By using \textit{ClipNet}, we expect to facilitate additional future research on data-driven vector graphics.
\section{ClipNet : Clipart collection of Man-made objects}
\label{sec:clipnet}
In this work, we introduce \textit{ClipNet} to facilitate the learning process of our clipart synthesis framework.
Inspired by available large datasets, such as ImageNet~\cite{imagenet_cvpr09} and ShapeNet~\cite{chang2015shapenet}, we intend to augment WordNet's structure by using clipart data.
Our primary goal was to compile a clipart collection in which each clipart contains the shape of the target category without any additional content.
Currently, \textit{ClipNet} contains 2D clipart of ten categories of man-made objects, such as chairs, tables, and airplanes.
\majorhl{
The reason for focusing on these ten categories of the man-made object is because they are commonly used during the animation, slides, and other 2D content creation processes.
}

\subsection{Data Characteristics}
During the data collection process, we have observed two crucial characteristics of clipart shared in online repositories: \ie~\textit{auxiliary content} and \textit{shading geometry}.
\begin{description}[style=unboxed,leftmargin=0cm]
\item[Auxiliary content]
A common characteristic of the clipart collected from online repositories is that it often contains additional auxiliary content apart from the shape representing the target category.
For example, the airplane clipart presented in \figname~\ref{fig:charac}(a) contains flow lines that indicate motions and clouds.
In the following paragraph, this airplane is considered the target shape and the remaining clipart is considered as the auxiliary contents.
\majorhl{The auxiliary content was removed because our goal was to enhance the ease of designing the vector shape of the target category, which is the most difficult part of clipart design.
As indicated in \mbox{\figname~\ref{fig:charac}}(a), the geometry of most of the auxiliary content was simpler than that of the target shape; thus, this content could be created easily with limited assistance.
}
\begin{figure}[t!]
\includegraphics[width=\linewidth]{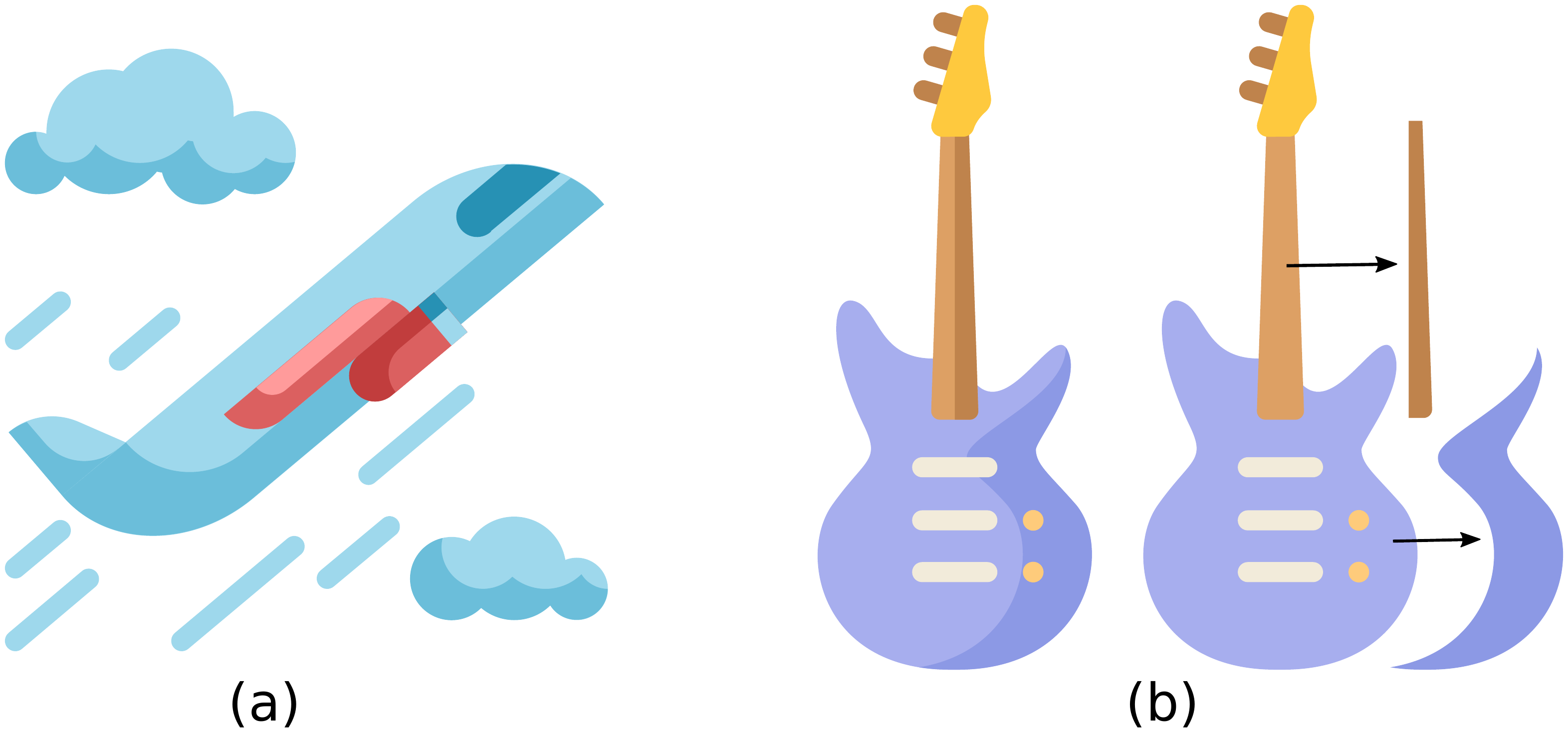}
\caption{Characteristics of the clipart obtained from online repositories:
(a) some clipart contain auxiliary content, such as clouds and motion lines, and 
(b) some path elements represent shading instead of geometry.
}
\label{fig:charac}
\end{figure}
\item[Shading geometry]
Another characteristic of the clipart is that it usually describes both \textit{geometry} and \textit{appearance} (\eg~color and shading)in the same file.
A 3D model file usually describes the geometry only, and its appearances are mostly defined in separate material or texture files.
However, the path elements in clipart sometimes represent the shading, \eg~reflection; (see \figname~\ref{fig:charac}(b)), which does not directly represent the shape.
Thus, we categorized the path elements presented in the collected clipart into two categories: the geometry path and the shading path. 
We labeled a path as a shading path if removing it did not change how people perceived a complete shape of the original category.
Then, we labeled the remaining paths as the geometry path.
For example, we labeled the two floating path elements in \figname~\ref{fig:charac}(b) as shading paths.
\majorhl{
In this work, we focused on generating the geometry paths that can form an object representing the target category under any lighting conditions; thus, we intended to remove the shading paths.
In addition, we regarded the shading geometry as the appearance part of the target shape, which can be treated as a separate problem to be addressed in the future.
}
\end{description}

\subsection{Data Preprocessing}
To obtain the target clipart without any undesired content, we performed two preprocessing steps. 
First, we identified all the auxiliary path elements and removed them from the input clipart.
Second, we identified the shading paths of the remaining paths of the first step.
We recruited five participants and asked them to perform the following two tasks.
\begin{description}[style=unboxed,leftmargin=0cm]
\item[Auxiliary  content removal]
Inspired by a previously proposed image segmentation annotation interface~\cite{bell2013opensurfaces}, we selected to use the polygon-based method, which allows the users to draw polygons in the region of the content-of-interest.
We first rendered the clipart into a raster image and recorded the pixels within each closed path.
For each clipart, we asked multiple users to draw polygons on the target shape and identified the corresponding path (see \figname~\ref{fig:anno_fig}(a)).
Because the polygons annotated by the participants were merely used for identifying relevant closed paths rather than representing the shape directly, the polygons were not required to be accurate.
For each clipart, we included the paths identified by all the users as the target shapes.

\item[Shading path removal]
The most direct approach to remove the shading path from the clipart is to ask users to remove it by using a vector graphics editor, such as Adobe Illustrator or Inkscape.
However, most users are not experienced in using such software; thus, we designed a simple annotation process that only required them to answer yes-no questions.
We removed the $i$-th path for each clipart with $N$ different paths and rendered the remaining paths into a raster image as $i$-th image.
In the interface, we presented the complete image and the $i$-th image side-by-side and asked the users to identify whether the shape in $i$-th image was still a complete shape of the target category (see the example pair and the question asked in \figname~\ref{fig:anno_fig}(b)).
If the user believed that the remaining paths in the $i$-th image represented the target category, we labeled the $i$-th path as the shading path.
The rationale behind the aforementioned task design was that the removed path was a shading path if it did not affect the user in perceiving the remaining paths as a complete object of the target category.
We asked the five participants to answer the question for each path and labeled a path as a shading path if all the users agreed that the remaining paths represented a complete object. 
\begin{figure}
\includegraphics[width=1.0\linewidth]{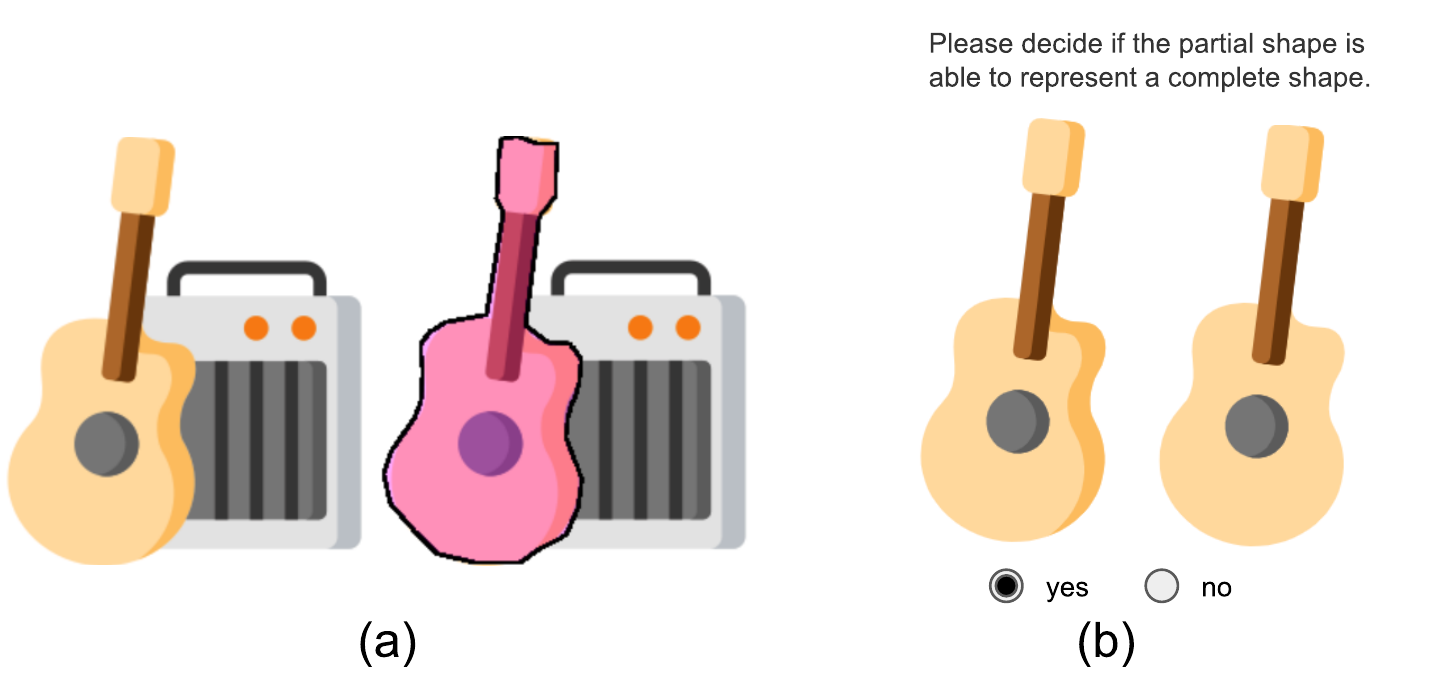}
\caption{Two annotation tasks for data preprocessing.
(a) \textit{Auxiliary content removal}: we asked users to draw polygons over the region of the target shape. 
(b) \textit{Shading path removal}: we presented a pair of shapes to the user and asked them to answer a ``yes-no'' question to identify the shading path.
In this example, the user answered ``yes,'' which indicates that even without the removed path, the remaining shape can still represent a complete \textit{guitar} shape; therefore, the removed path was regarded as a shading path.
}
\label{fig:anno_fig}
\end{figure}
\end{description}

The processed clipart was compiled into the \textit{ClipNet} dataset, which was used to train different modules that were adopted in the proposed method.
\section{Problem Formulation}
\label{sec:prob_form}
\begin{figure}[t!]
\includegraphics[width=\linewidth]{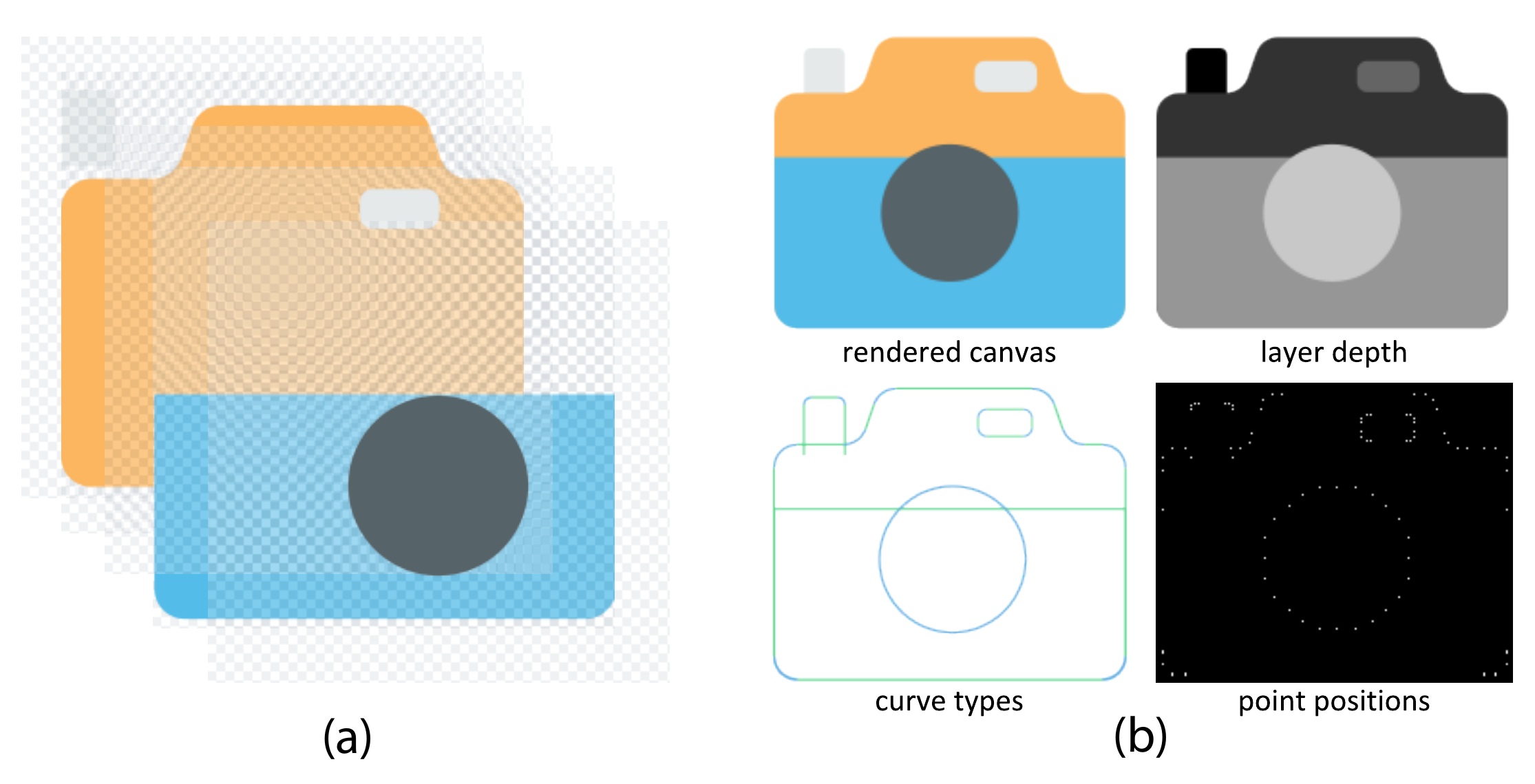}
\caption{
\majorhl{
(a) Layer representation of a vector clipart image.
(b) A multichannel image including various visual representations of the current canvas.
}
}
\label{fig:layer_viz}
\end{figure}
This work aimed to synthesize clipart comprising a multilayer vector representation (see \figname~\ref{fig:layer_viz}(a)) conditioned on a selected category (\eg~airplanes, chairs, or cameras) and a target image.
Each layer $i$ contains a closed path $S_i$, and it's filled RGB color $\phi_i$.
\majorhl{
This assumption is commonly used for artist-drawn clipart shared on the Internet and thus has also been adopted by previous clipart vectorization works~\mbox{\cite{FLB17}}.
}
To render each layer, we generated a binary mask $M_i$ (with a value of $1$ inside the closed path and $0$ outside) at the target resolution.
The $n$-th layer of the rendered image of this clipart can be formulated as follows:
\begin{align}
I^n = I^{n-1} \cdot (1-M_i) + \phi_i \cdot M_i.
\label{eq:composite}
\end{align}
The symbol $I_i$ represents the rendered image of the content in layer $i$, and $I^{i}$ represents the rendered image of the content in layers 0 to $i$.

This paper focuses on learning generative models of clipart of man-made objects.
We formulated the synthesis process as a sequential prediction problem.
Our synthesis model, whose main building blocks comprise deep neural networks, can encode complex relationships between parts of the clipart.

\section{Synthesis Model}

\begin{figure*}[t!]
\includegraphics[width=\linewidth]{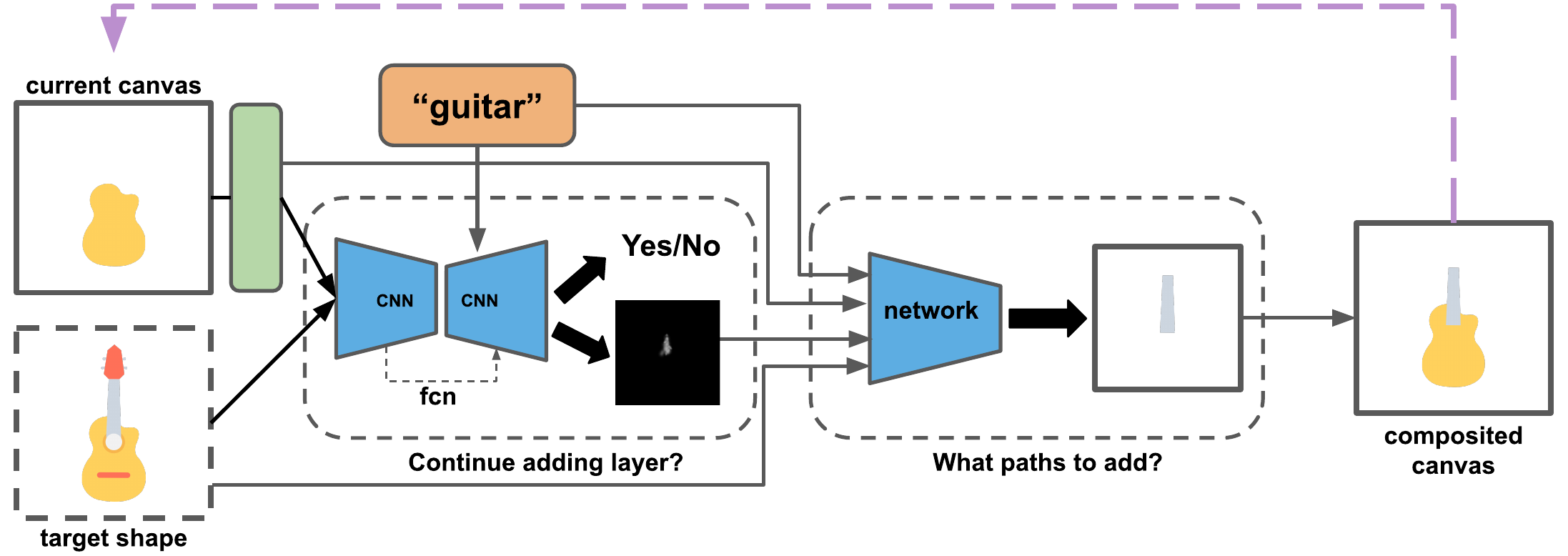}
\caption{The pipeline of our synthesis model.
For each layer $n$, we take the synthesized shape until layer $n-1$ and obtain it s visual representation, including the rendered image $I^{n-1}$ and the curve information (through the $V$ function in the green box).
The input of the first module of the proposed synthesis model is a visual representation of the current canvas, chosen category (guitar in this figure), and target shape (optional).
The first module decides whether to continue adding a new layer onto the current canvas and predict the probability map of the potential path.
If the model decides to add a new layer, we use the same set of input combines the predicted probability map of the new path to predict a sequence of path elements and its filled color.
We composite the latest predicted layer with the existing layers and use them as input to the synthesis model iteratively.
}
\vspace{-2.5mm}
\label{fig:generative}
\end{figure*}
\begin{figure}[t!]
\includegraphics[width=\linewidth]{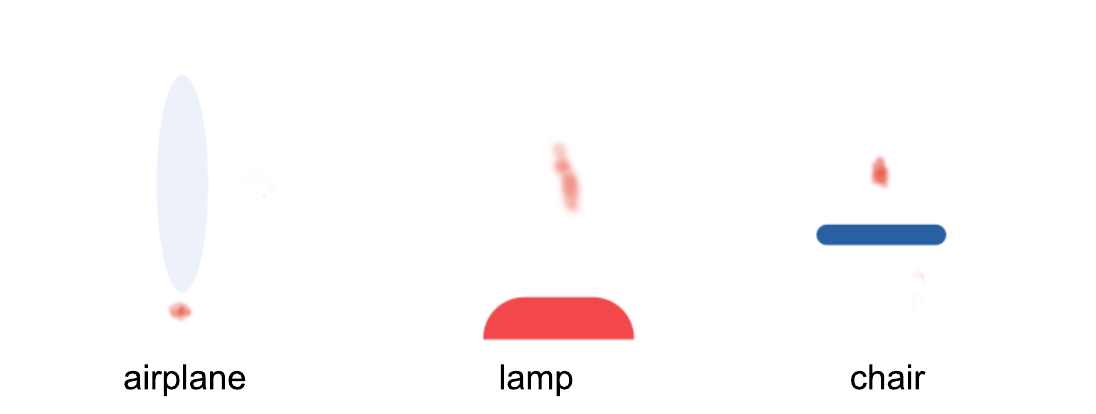}
\caption{
The predicted probability maps of the center position of the closed path in the next layer.
}
\vspace{-2.5mm}
\label{fig:cont_pred}
\end{figure}
Our vector clipart synthesis model has two inputs: a target category and a target raster image, whose source depends on the application.
For example, for vectorizing an existing vector clipart image, we can use a low-resolution raster clipart image as the target raster image.
To synthesize novel clipart belonging to a specific category, we trained a generative model for synthesizing a novel raster clipart image as the target image (refer to \figname~\ref{fig:first_syn} and \ref{fig:second_syn} for examples).
We can synthesize diverse novel clipart by leveraging the diversity of the synthesized novel raster clipart image.

\figname~\ref{fig:generative} illustrates our synthesis process.
Because a vector clipart image of a man-made object is usually composed of several separate parts placed in layers, our model synthesizes the entire clipart iteratively (\ie~synthesizes each layer sequentially).
After synthesizing the $(n-1)$-th layer, our model renders the constructed clipart ($S^{n-1}$, where $S^i$ denotes the shapes in layers 0 to $i$) into rendered clipart image ($I^{n-1}$) and attaches additional information to it to form a visual representation ($V(S^{n-1})$) (described in \secname~\ref{sec:vis_rep}).
Let us take the visual representation as the input. 
The first step is to predict whether our model should synthesize a new layer.
If the model decides to predict the next layer, the second step of our model is to predict a closed path, including the positions of all the control points (including the start and end points), and the filled RGB colors ($\phi_{n}$).
After these two steps, our model synthesizes a new layer composed of the latest predicted path and its filled color.
We obtain the latest image of this layer ($I^n$) by applying \eqname~\ref{eq:composite} and re-iterate the synthesis process by using the latest synthesized clipart.

\subsection{Visual representation of the canvas}
\label{sec:vis_rep}
Unlike a raster image, which is only described by pixel values, vector clipart is also described by different layers, various curves, and the relationships between curves.
To obtain a unified representation of the information encoded in the clipart, we converted the aforementioned data into multiple 2D images and stacked all of them into a multichannel image.
Then, we used a deep convolutional network as the feature extractor to describe the clipart.

The multichannel image included the following components (see \figname~\ref{fig:layer_viz}(b)):
\begin{description}[leftmargin=0cm]
\item[Rendered canvas:] The rendered canvas is the rendered image of the current canvas.
\item[Curve types:] We rendered different curves according to their types.
\item[Layer depth:] We rendered each closed path according to its layer depth.
\item[Point positions:] We highlighted the pixels that contained the control points of the existing curves.
\end{description}
\subsection{First step: Decision on whether to continue adding layers?}
\label{sec:cont_pred}
The first step of our synthesis model is to decide whether to continue adding new layers to the canvas.
This function takes the current canvas observations as input and outputs (i) the probability of continually adding a new layer and (ii) the probability map of the center location of the next closed path.
The observations of the current canvas include the components described in the following text.
The first component is a feature extracted from the visual representation ($V$) of the current canvas state.
We fine-tuned a pretrained deep convolutional network (we use ResNet-50) as the feature extractor.
The second component is the one-hot encoding of the target category ($t$) to be synthesized.
The information about this category provides useful context for the canvas; \ie~a similar path represents different meanings under different object categories.
The third component is the count of the existing layers (same as the number of closed paths) already synthesized on the canvas.
The fourth component is the rendered raster image of the target shape ($I_T$).
We illustrate the observations in \figname~\ref{fig:generative}.

\subsubsection{Training data and process}
\begin{description}[style=unboxed,leftmargin=0cm]
\item[Training data] We used the \textit{ClipNet} dataset to train our model.
We built a dataset with 50\% positive and 50\% negative examples.
We created positive examples (\ie~examples in which the ``continue to add path'' prediction is made by the model) by using the following three approaches: 
\begin{enumerate}
    \item Selection of one complete clipart and the removal of the last path from the original clipart: We generated a location probability map by using the center location of the last path being removed.
    \item Selection of one blank clipart: We generated its corresponding location probability map by using the center location of the first path.
    \item Selection of one complete clipart and the removal of a random number of paths: We generated a location probability map by using the center location of the path with the largest depth value (i.e., the deepest path).
\end{enumerate}
In the aforementioned three approaches of generating positive examples, the synthesis order is naturally encoded through the location probability map (because we used the center location of the next path as the ground-truth location probability map). 
By contrast, the negative examples (\ie~examples in which “do not continue to add paths” is predicted by the model) were developed by directly using a complete clipart image.
\end{description}
In total, we create 4000 examples for training the first module.

We trained the first module by using a fully-convolutional encoder-decoder
network (FCN~\cite{long2015fully}) to predict both parts simultaneously (refer to \figname~18 in the Supplemental Material for the architecture).
To train the location probability part, we used pixel-wise cross-entropy loss.
To train the ``continue synthesis part'', we used two additional linear layers after the decoder output (\ie~the probability map) to obtain a binary prediction.
We trained this part of the model as a binary classification problem by using the binary cross-entropy loss.
We trained this module by using an Adam optimizer~\cite{kingma2014adam} for 200 epochs with a learning rate of $0.001$.
During inference, this model terminated the entire synthesis process if the predicted probability was less than $0.5$.
\figname~\ref{fig:cont_pred} shows several examples from different categories as well as their predicted probabilities.
\subsection{Second step: decision on which path to add next?}
In the second step, we aimed to generate a new layer that contained one closed path (composed of several connected curves and an RGB color).
We aimed to determine a function $F^{\beta}$ that takes the observation of the current canvas as input and outputs the sequential curve predictions (including the curve type and control point positions).
\majorhl{
In this work, we used two types of curves: a line curve and cubic B{\'e}zier curve.
}
The final action involved predicting an RGB color $\phi$ that fills the closed path.
We used four types of observations in our prediction model.
The first two observations were the visual representation ($V$) of the current canvas and the one-hot encoding of the target category $t$.
The third observation was the center location probability map $I_p$ output in the first step (refer to in \figname~\ref{fig:cont_pred} for examples).
The last input was the target image $I_T$.
The generative model was formulated as a function $F^{\beta}$ as follow:
\begin{align}
[C_1, C_2, ... , C_K, \phi] = F^{\beta}(V, t, I_p, I^T),
\end{align}
\majorhl{
where $C_1$ to $C_K$ indicate the predicted curves of the closed path, including the curve type and control point positions.
}

\subsubsection{Network architecture}
\begin{figure}[t!]
\includegraphics[width=0.95\linewidth]{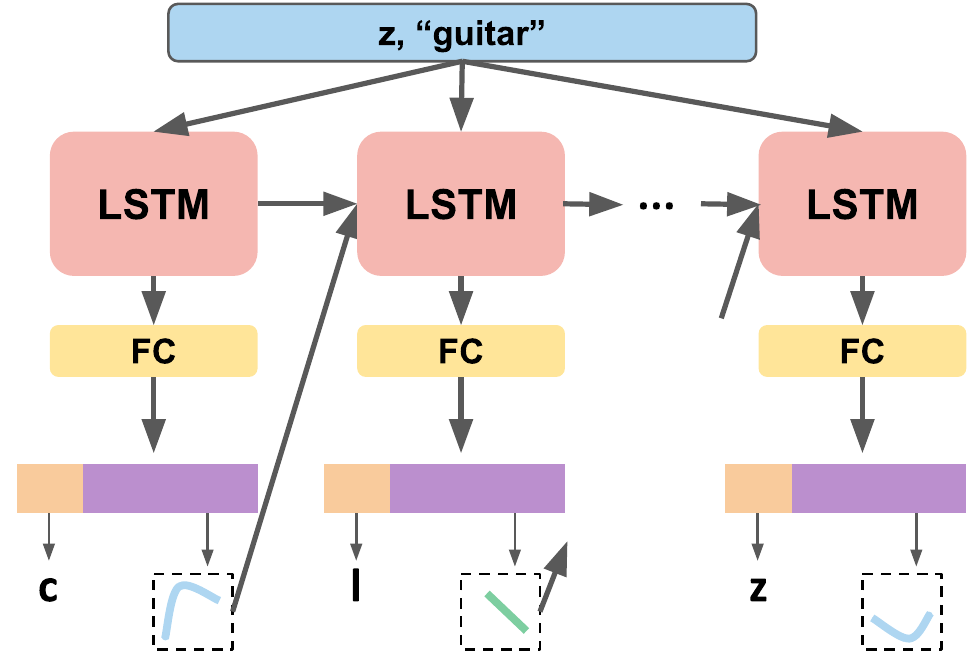}
\caption{
Network architecture of the recurrent decoder.
At each step, the LSTM architecture receives the input, including the latent vector of the current canvas ($z$),  
the previous timestep's output (path element), and the chosen category (``guitar'').
Then, it generates the type, start point, end point, and control points of the new path.
}
\vspace{-2.5mm}
\label{fig:decoder}
\end{figure}

We implemented a generative function $F^{\beta}$ with a Recurrent Neural Network (RNN) decoder.
We used this architecture because RNNs have demonstrated high accuracy in tasks related to sequential and time-series data processing.
\figname~\ref{fig:decoder} presents the detail of our framework for predicting a new path and color in the new layer.
By considering the aforementioned four observations mentioned above, the decoder sequentially predicts the type of each path (\ie~line (``l''), cubic B\'{e}zier curve (``c''), or end of the curve (``z'')) and its control point positions (including the start and end points).

\begin{description}[style=unboxed,leftmargin=0cm]
\item[Visual encoder]
For each training example, we stacked the visual representation ($V$), the center probability map ($I_p$), and the target image ($I_T$) by using Resnet50~\cite{he2016deep} as our visual feature extractor.
The output feature vector obtained from Resnet50 was concatenated with the one-hot target category vector and passed to two linear layers (with 128 and 64 neurons, respectively).
The output was a 64-d latent vector ($z$), which was passed to the path generator to predict a sequence of curves.
\end{description}

\begin{description}[style=unboxed,leftmargin=0cm]
\item[Path generator]
In the path generator, we first took the latent vector $z$ as the input and used a fully-connected layer to predict the path's starting position.
We used Long Short-Term Memory (LSTM)~\cite{hochreiter1997long} as the recurrent architecture for our sequential generator.
At each time step $i$ of the decoder RNN, we feed the previously predicted control points, the latent vector $z$, and one-hot encoding of the target object category $t$ and concatenated them into an input vector $x_i$.
The outputs of each step were (i) the type of the curve and (ii) the start point, end point, and control points.
\figname~\ref{fig:decoder} presents the details of the recurrent path generator.
\end{description}

\subsubsection{Loss Function}
\label{sec:loss}
\begin{figure}[t!]
\includegraphics[width=\linewidth]{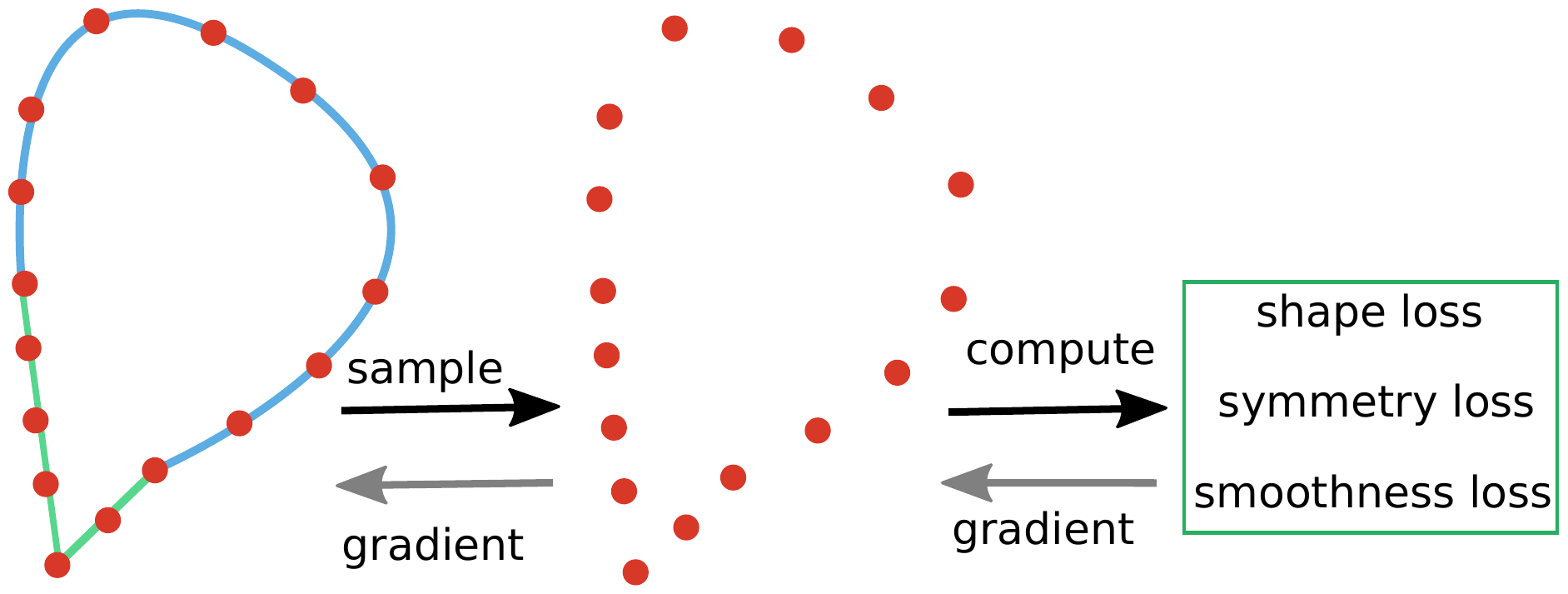}
\caption{Sampling on the predicted path, which enables end-to-end training.
The gradients of the sample points are backpropagated onto the predicted control points. 
}
\vspace{-2.5mm}
\label{fig:sampling}
\end{figure}

We used the predicted path elements to construct a closed path ${S}'_i = \{C_1, C_2, ... , C_K\}$ for a new layer $i$.
To train our generative models, we defined a loss function composed of geometric and rendering losses as follows:
\begin{align}
L_{total} = L_{geom} + L_{render},
\end{align}
where the geometric loss is expressed as follows:
\begin{align}
L_{geom} = (L_{chm} + L_{mover}) &+ \omega_{sym} (L_{sym}+L_{csym}) \\&+  \omega_{smooth}L_{smooth}.
\end{align}
\begin{description}[style=unboxed,leftmargin=0cm]
\item[Curve Sampling]
To measure the similarity between the predicted path ${S}'_i$ and target path $S_i^{T}$, we computed the shape loss (both $L_{chm}$ and $L_{mover}$), symmetry loss ($L_{sym}$), and smoothness loss ($L_{smooth}$) on a set of points sampled on both paths.
As illustrated in \figname~\ref{fig:sampling}, we used a uniform sampling strategy on each curve $C$; that is, if $p$ is a sample point on a curve $C$, its position can be represented as follows:
\begin{itemize}
    \item \textbf{line}.  $p = (1-t) \cdot p^i_0 + t \cdot p^i_1$, where $p^i_0$ and $p^i_1$ are the start and end points of the line $i$, respectively, and $t \in [0,1]$.
    \item \textbf{cubic B{\' e}zier curve}. $p = (1-t)^3 \cdot p_0^i + t(1-t)^^2 \cdot p_1^i + t^2 (1-t) \cdot p_2^i + t^3 \cdot p_3^i$, where $p_0^i$ and $p_3^i$ are the start and end points, respectively; $p_1^i$ and $p_2^i$ are the two control points of path element $i$; and $t \in [0,1]$.
\end{itemize}
We sampled $n$ sample points ($n=200$ in this paper) for the entire path $S_i$ and uniformly distributed them to each path element (\ie~if there are five path elements in $S_i$, we uniformly sample $n/5$ sample points on each of them).
Because the sampled points were represented using the predicted control points, they were differentiable throughout the entire learning process.
\end{description}

\begin{description}[style=unboxed,leftmargin=0cm]
\item[Shape Loss]
We used the ordered Chamfer loss ($L_{chm}$) and Earth Mover's distance loss ($L_{mover}$) to measure the similarity between the predicted path ${S}'_i$ and the target path $S_i^T$.
On the basis of the 2D sample point set $\mathbf{P}_{C}$ on the predicted path and the $\mathbf{P}_{C}^T$ on the target path, we computed the following losses:
\end{description}
\begin{itemize}
\item \textbf{Ordered Chamfer loss}
The traditional Chamfer loss used in~\cite{fan2017point} assumes the existence of an unordered point set (permutation invariant); however, in this work, the sampled points on the predicted path exhibited well-defined ordering.
To encourage the predicted path to follow the ordering of the target path, 
we defined our matching loss as follows:
\begin{align}
L_{match}(\mathbf{P}_{C}, \mathbf{P}_{C}^T) = \min_{j \in [0, ..., K-1]} \sum_{i=0}^{K-1} \| p_i - p^T_{(j+i)\%K} \|^2_2 \\
L_{chm}(\mathbf{P}_{C}, \mathbf{P}_{C}^T) = L_{match}(\mathbf{P}_{C}, \mathbf{P}_{C}^T) + L_{match}(\mathbf{P}_{C}^T, \mathbf{P}_{C})
\end{align}
where $K$ is the number of sample points, and all points on a path are sampled in the same order (clockwise).
\item \textbf{Earth Mover's loss}
The earth mover's distance loss can be computed as follows:
\begin{align}
L_{mover}(\mathbf{P}_{C}, \mathbf{P}_{C}^T) = \min_{\Phi:\mathbf{P}_{C} \rightarrow \mathbf{P}_{C}^T} \sum_{p \in \mathbf{P}_{C}} \| p - \Phi(p) \|_2^2,
\end{align}
where $\Phi:\mathbf{P}_{C} \rightarrow \mathbf{P}_{C}^T$ is a bijection.
\end{itemize}
\begin{description}[style=unboxed,leftmargin=0cm]
\item[Symmetry Loss]
If a target path exhibits symmetry, then we want the predicted path to preserve this symmetry property.
However, since the topology of the path is not always guaranteed to be symmetric (\ie~the curves in a path do not always have symmetric control points even if the sample points on the curves are symmetric).
This information is crucial because we desire to equip the predicted path with editability.
To address the requirements for editability and symmetry, we designed two losses for symmetry loss: sample point symmetry loss and control point symmetry loss.
\end{description}
\begin{itemize}
\item \textbf{Sample point symmetry loss}
We sampled a point set $\mathbf{p}'_{C}$ on the mirrored predicted path and computed both the Chamfer loss and the earth mover's distance loss between the mirrored point set and the point set sampled from the target path as follows:
\begin{align}
L_{sym}(\mathbf{P}_{C}, \mathbf{P}_{C}^T)= L_{chm}(\mathbf{P}'_{C}, \mathbf{P}_{C}^T) + L_{mover}(\mathbf{P}'_{C}, \mathbf{P}_{C}^T)
\end{align}
\item \textbf{Control point symmetry loss}
By using the predicted control points $\mathbf{Q}$, we first created a mirrored control points ${\mathbf{Q}}'$ and then compute the Chamfer distance between predicted and mirrored controlled points as follows:
\begin{align}
L_{csym}= L_{chm}({\mathbf{Q}}', \mathbf{Q}).
\end{align}
We computed the loss using the following control points:
\begin{itemize}
        \item \textbf{line}: $\mathbf{Q} = \{ p_{\text{start}}, p_{\text{end}}\}$
        \item \textbf{cubic B\'{e}zier curve}: $\mathbf{Q} = \{ p_{\text{start}}, p_{\text{control1}}, p_{\text{control2}}, p_{\text{end}}\}$
\end{itemize}
\end{itemize}
\begin{description}[style=unboxed,leftmargin=0cm]
\item[Smoothness Loss]
To preserve the local geometric details, including the sharp corner and smooth part of the target path, we expected the Laplacian coordinate~\cite{sorkine2004laplacian} of the sample point on the predicted path ${S}'_i$ to be the same as that on the target path $S_i^T$. 
We defined the smoothness loss as follows:
\begin{align}
L_{smooth} = \sum_i (\mathcal{L}(p_i)-\mathcal{L}(p_i^T)),
\end{align}
where $p_i \in \mathbf{P}_C$ and $p_i^T \in \mathbf{P}_C^T$, and 
\begin{align}
\mathcal{L}(p_i) = p_i - \frac{1}{2}(p_{i+1}+p_{i-1})
\end{align}
is the Laplacian coordinate defined on the polyline formed by the sample points on the path.
\end{description}


\begin{description}[style=unboxed,leftmargin=0cm]
\item[Rendering Accuracy Loss]
To better predict the color and refine the path geometry, we leveraged a novel differentiable renderer for vector graphics~\cite{Li:2020:DVG}.
This differentiable rendering loss aims to propagate the changes in the image observation to the vector graphics parameters.
The vector graphics differentiable renderer supports different SVG primitives such as paths (with quadratic or cubic segments), lines, and circles, and each path can have a fill color and stroke color.
In this work, we only used cubic curves and lines and filled color for closed paths.

To illustrate the concept and the ability of this differentiable vector graphics renderer, let us first assume we have a curve set $\mathbf{C}$ with their corresponding color properties $\mathbf{\Psi}$.
The rendering process can be described using a function $\varphi$: $I = \varphi(\mathbf{C}, \mathbf{\Psi})$.
The rendering accuracy loss can be defined as follows:
\begin{align}
L_{render} = \|(I-I^T)\|^2,
\end{align}
where $I^T$ is the target image during training.
The determination of the optimal curve parameters $(\mathbf{C}^{*},\mathbf{\Psi}^{*})$ can be formulated as an optimization problem as follows:
\begin{align}
(\mathbf{C}^{*},\mathbf{\Psi}^{*}) = \underset{\mathbf{C},\mathbf{\Psi}}{\mathrm{argmin}} \quad L_{render},
\label{eq:diff_opt}
\end{align}
We aimed to determine an optimal set of new positions ($\mathbf{C}^{*}$) and color properties ($\mathbf{\Psi}^{*}$) of the initial curve set and color properties so that the rendered image $I$ matched the target image $I^T$.
\majorhl{
For the derivatives derived in~\mbox{\cite{Li:2020:DVG}}, both curve set positions and color properties are treated as continuous parameters; thus, we can optimize these parameters together by using the gradient-based optimization method, as indicated in \mbox{\figname~\ref{fig:diffvg}}.
}
The differentiable vector graphics renderer is not the main contribution of this paper but was proposed by another work~\cite{Li:2020:DVG}.
The novelty of this work lies in facilitating the clipart synthesis process by using this differentiable renderer.
\begin{figure*}
\includegraphics[width=\linewidth]{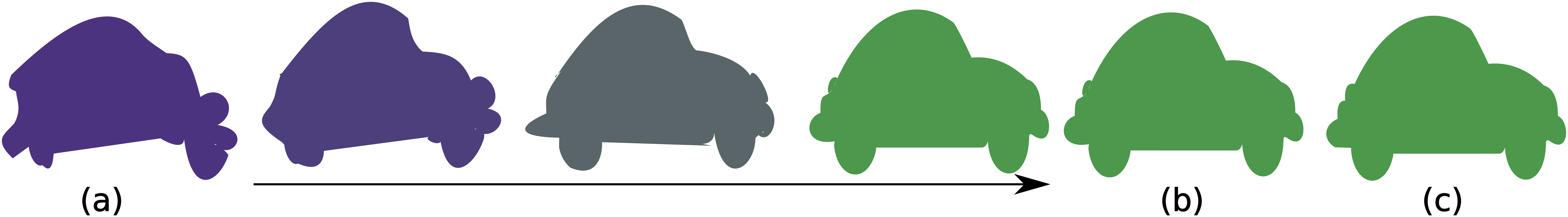}
\caption{(a) Randomly manipulation of the initial position of the control points and the initial filled color. By optimizing \eqname~\ref{eq:diff_opt} with the target image (c), we obtained the intermediate shape and color.
Both the shape and the filled color were gradually updated toward those of the target image. 
}
\label{fig:diffvg}
\end{figure*}
\end{description}

\subsubsection{Training data and process}
\label{sec:mod2_data}
\begin{figure}
\centering
\includegraphics[width=\linewidth]{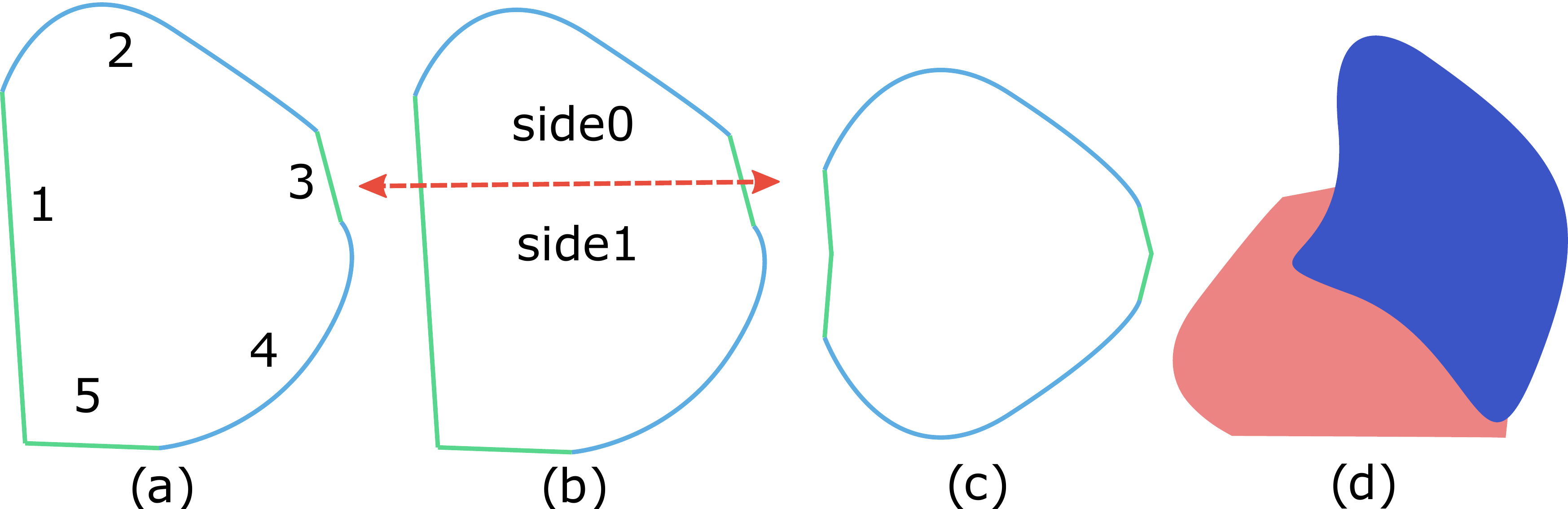}
\caption{
Random path generation process.
(a) In this example, we synthesized a closed path by using five curves.
The numbers indicate the synthesis order.
(b) If symmetry was desired, we selected a symmetry axis (the red dot line) and picked a side that we wished to preserve.
(c) We selected ``side0'' of the path and mirrored the curve on this side to generate the final path.
(d) We also generated multiple random paths with different colors as the training set.
}
\label{fig:rand_path}
\end{figure}
\begin{description}[style=unboxed,leftmargin=0cm]
\item[Training data]
To train our path generator, we leveraged two training datasets: a category-specific dataset collected in \textit{ClipNet} and a random synthesis set, which was formed by augmenting \textit{ClipNet} with a non-category-specific path set generated by a random closed-path generator. 
We synthesized a random closed path by first deciding the curve count in the path and then determining whether this path was symmetric.
With an empty canvas, we progressively generated a path without self-intersection.
First, we randomly selected a curve type (line or cubic B\'{e}zier curve) and randomly decided the positions of all the start, end, and control points.
Then, we repeated the aforementioned procedure and validated whether the current curve intersected with (i) itself and (ii) all the generated curves.
If we considered a path as symmetric, we randomly selected a symmetric axis, and we use this axis to separate the generated path into two parts. 
We randomly selected one of the two sides, and mirrored the path's selected side, and removed the unselected part.
The synthesis process is illustrated in \figname~\ref{fig:rand_path}.
\item[Curriculum training process]
Inspired by the curriculum learning training strategy used in \cite{bengio2009curriculum,hacohen2019power}, 
we categorized the entire training process of our generative model into three stages:
\begin{enumerate}
    \item We used the paths in a random synthetic set to train the path generator. 
    For each target shape, only a single path existed (\ie~the generator was only required to focus on generating a single visible path).
    \item We generated multiple paths on the canvas and used the one on the top as the path to be generated (refer to \figname~\ref{fig:rand_path}(d) for an example).
    In this stage, we encouraged the generator to recognize that some content in the target shape was generated in the previous step.
    \item We used the \textit{ClipNet} path data to train the path generator to predict a category-specific path.
\end{enumerate}
In the first two stages, we used the data without category information to train the generator.
The aim of this action was to encourage the generator to learn how to generate single and multiple paths first and then learn about their relationships conditioned on the category.
We trained the first stage for 200, 200, and 300 epochs in the first, second, and third (final) stages, respectively.
The curriculum training process was conducted to encourage the path generator to familiarize itself with the most straightforward task first (\ie~to generate the non-category-specific curve geometry), to learn how to predict color, and finally to learn the correlation between the generated curve and the categories.
\end{description}

\subsection{Shape Regularization}
\begin{figure}
\centering
\includegraphics[width=0.8\linewidth]{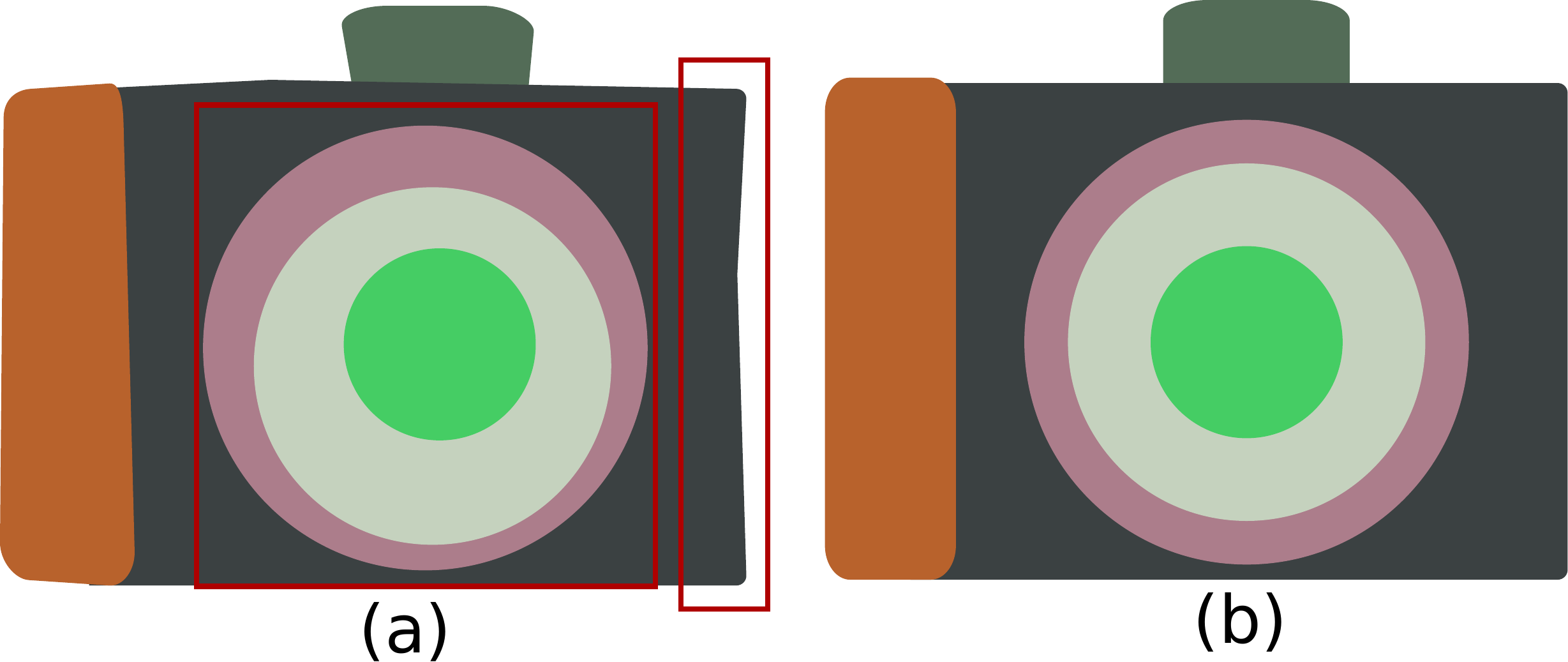}
\caption{(a) Synthesized clipart before regularization. 
Although the predicted paths clearly represent the desired category (\ie~a camera), several parts can be further regularized into superior clipart (\ie~the parts inside the red rectangles).
(b) Resultant clipart after applying axis-aligned lines, concentric, parallel lines, and arc-like regularization to improve the synthesized clipart into a more regularized clipart.
}
\label{fig:regular_path}
\end{figure}

After predicting all the layers, we regularized the synthesized paths 
by identifying paths with similar but not identical characteristics, such as axis-aligned lines, concentric curves, parallel lines, and closed paths with similar shapes, and forced them to be identical (see \figname~\ref{fig:regular_path}).
Given an identified group of a closed paths, our method progressively enforces the regularities in the following order (similar to \cite{globfit,hoshyari2018perception}):
\begin{description}[style=unboxed,leftmargin=0cm]
\item[Axis-aligned lines]
If several line segments in the closed paths are close to the horizontal or vertical axis \majorhl{(\mbox{\ie}~the angle between the lines
and the vertical or horizontal axis less than \mbox{$10\degree$})}, we force them to be axis-aligned.
\item[Arc-like path]
If the closed path is very close to a circle, our method can fit the closest circle to replace the predicted closed path.
\majorhl{
We detected an arc-like path by first sampling points on the path and computed the average distance between these sample points and the closest arc.
If the average distance is less than 0.5, we regarded this path as an arc.
}
\item[Concentric path]
If the shapes of several closed paths are very close to a circle, and their centers are very close to each other, our method can enforce them to center at the same position.
\majorhl{
We treated two paths as co-centric if the distance between their centers is less than 1/10 of the longest side of one of their bounding boxes.
}
\item[Parallel lines]
We sample line segments across different closed paths and determine if they can be forced as parallel.
\majorhl{
We treated lines as parallel if their angle of directions between each other is less than \mbox{$10\degree$}.
}
\end{description}

\section{Results and Evaluations}
\begin{figure}[t!]
\includegraphics[width=\linewidth]{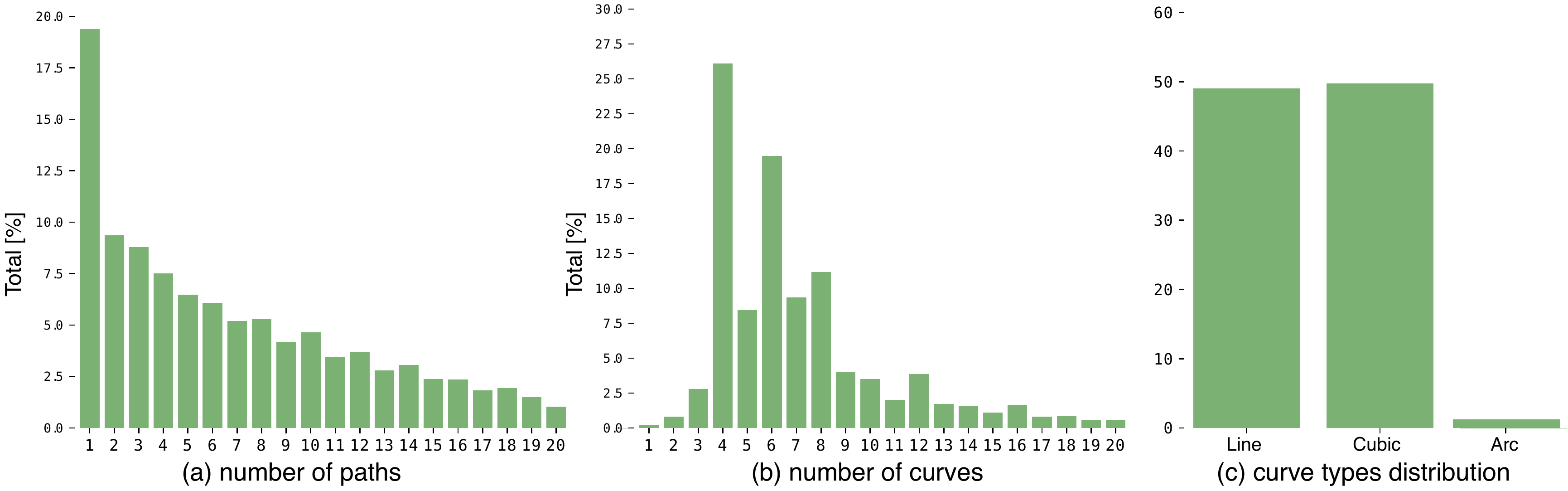}
\caption{
Each clipart in our dataset comprises multiple closed paths, each of which is composed of numerous curves.
(a) Distribution of the number of paths, (b) the distribution of the number of curves of each path, and (c) the distribution of the curve types of the entire dataset.
}
\vspace{-2.5mm}
\label{fig:data_stats}
\end{figure}
\subsection{Data analysis}
In this paper, we introduced a new clipart dataset called \textit{ClipNet}, which currently contains 2000 clipart of man-made objects belonging to ten categories (please refer to the Supplemental Material for additional information on the clipart of each category in \textit{ClipNet}.
Each clipart in our dataset is composed of multiple paths, each of which is composed of multiple curves.
To indicate the complexity of the clipart in our dataset, some statistics are illustrated in \figname~\ref{fig:data_stats}.
As indicated in \figname~\ref{fig:data_stats}(a), most of the clipart contains less than ten separate paths, and most of these paths only contains less than eight separate curves (\figname~\ref{fig:data_stats}(b)).
The distribution of the types of curves in the current dataset is presented in \figname~\ref{fig:data_stats}(c).
As observed in \figname~\ref{fig:data_stats}(c), line and cubic B\'{e}zier curves are the two dominant types of curves in the dataset; therefore, we modeled these two types of curves in our work.



\subsection{Implementation details}
We implemented our synthesis model in PyTorch~\cite{paszke2017automatic}.
We used a feature extractor (Resnet50 architecture~\cite{He2016}) and a layernorm~\cite{Ba2016LayerN} LSTM as our recurrent decoder.
We conducted all the experiments on a desktop computer with an Intel Core i7-7800 CPU (3.5GHz) and a GeForce GTX 1080 Ti GPU (11GB memory).

\subsection{Ablation study}
We conducted a set of experiments to validate our synthesis model design's efficacy and the proposed loss functions.
For the validation, we used our random path generator to generate different training and testing datasets.
\subsubsection{Effect of the Recurrent Decoder}
We modeled our curve predictor by using a recurrent decoder.
Two reasons existed for adopting this approach.
First, we did not know the count and the types of curves that formed the desired path in advance.
Second, even if the curve count is known, the recurrent structure can still provide superior results in the sequential synthesis of the curve better because at each time step, the recurrent model feeds the previously predicted result so that the model can predict the next positions conditioned on the existing predictions~\cite{Yin2017ComparativeSO}.

To validate the advantage of the recurrent model, we conducted a restricted experiment.
We used the random path generator described in~\secname~\ref{sec:mod2_data} to generate paths with four cubic B{\' e}zier curves.
This step was performed to eliminate the first problem of the CNN architecture and focus on the second problem (\ie~the advantage of the recurrent architecture).
We generated 10000 random paths with random colors and rendered them into $64\times64$ raster images.
Of the 10000 generated images, 9000 formed the training set and 1000 formed the test set.
In both methods, we used the same image encoder (Resnet50) but different decoders.
\majorhl{
For a fair comparison, we designed a recurrent CNN architecture.
In each step, the network predicts the current 2D position using the concatenated feature of the extracted feature and the previous 2D position.
Please find the architecture in \mbox{\figname~19} in the supplemental material).
}

\begin{figure}[t!]
\includegraphics[width=\linewidth]{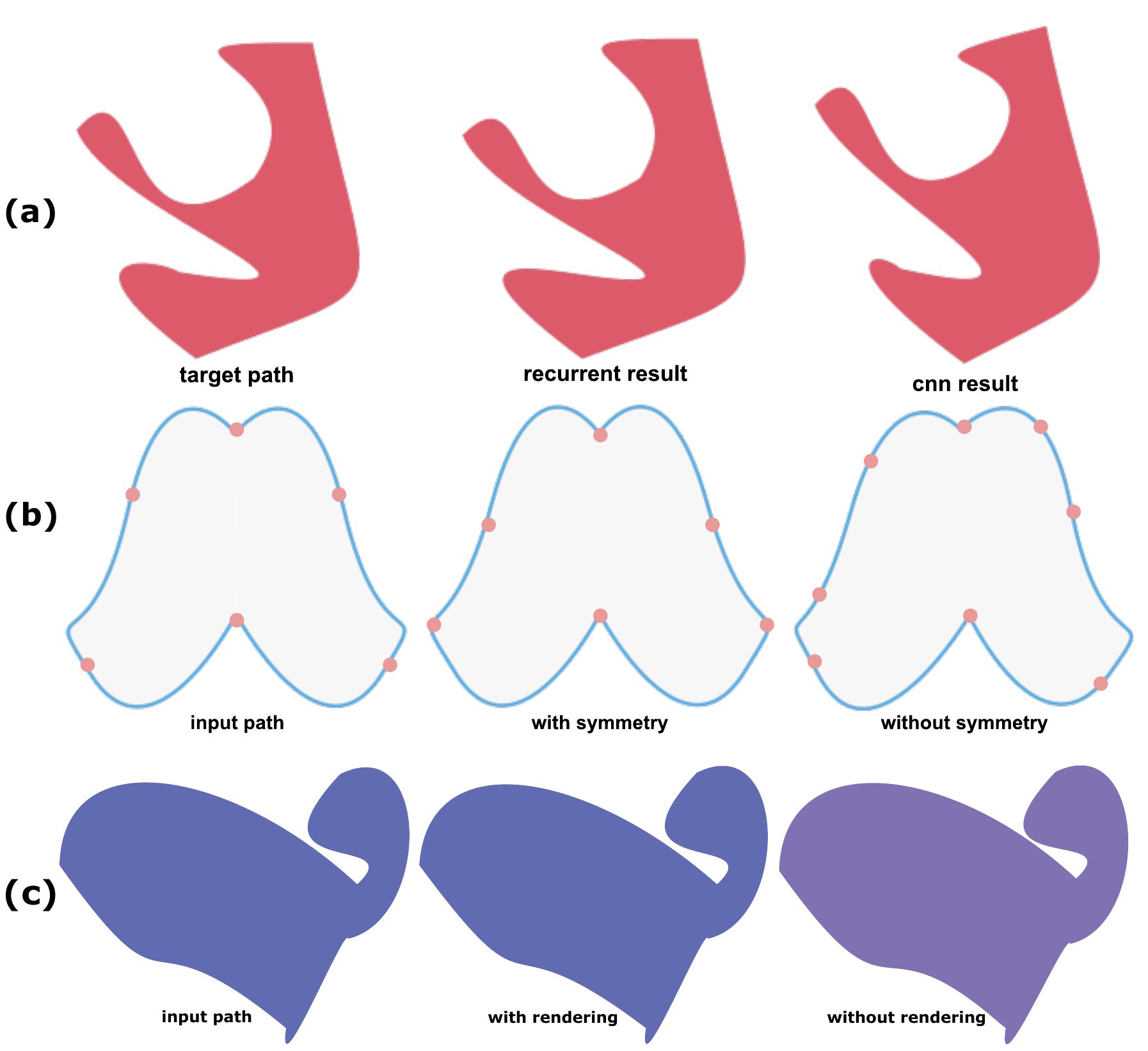}
\caption{
Example results obtained in the ablation study. 
}
\vspace{-2.5mm}
\label{fig:ablation}
\end{figure}

We used a joint loss function that includes the shape, symmetry, and smoothness losses.
\figname~\ref{fig:ablation}(a) presents examples of paths reconstructed using the two methods (only the curve geometry was predicted, whereas the color was assigned randomly during path generation.)
\begin{description}[style=unboxed,leftmargin=0cm]
\item[Quantitative metrics]
We evaluated the quality of the two methods by computing the L2 image distance and Chamfer distance.
For computing the image distance, we rendered the predicted curve and computed the image difference between the rendered image and the input images.
\end{description}
\begin{table}
\begin{center}
\begin{tabular}{ c| c c } 
 \Xhline{1.2pt}
  & recurrent model & cnn model \\ \Xhline{1.2pt}
 L2 image difference & 24.622 & 55.331 \\ 
 Chamfer distance & 1.11 & 12.69 \\ 
 \Xhline{1.2pt}
\end{tabular}
\caption{Quantitative results obtained for the test set by using the recurrent decoder and non-recurrent CNN-based predictor.
We present the average image difference and Chamfer distance on the test set.
}
\label{tab:recurrent}
\end{center}
\end{table}
For calculating the Chamfer distance, we sampled 200 points on both the input and predicted curves and then computed the Chamfer distance between these two sets of sample points.
\tabname~\ref{tab:recurrent} lists the quality statistics, indicating that the recurrent model performs better in terms of both metrics.

\subsubsection{Effects of symmetry loss}
In the proposed method, two types of symmetry loss are examed: sample point symmetry loss and control point symmetry loss.
To effectively evaluate these symmetry losses effectively, we generated 10000 random paths with symmetry and rendered them into $64\times64$ raster images. 
Of the 10000 generated images, 9000 formed the training set and 1000 formed the test set.
\figname~\ref{fig:ablation}(b) indicates the paths synthesized by the models with and without the symmetry loss.
\begin{description}[style=unboxed,leftmargin=0cm]
\item[Quantitative metrics]
We used the symmetry axes generated during the random path generation to
compute both loss functions: $L_{sym}$ and $L_{csym}$ (we sampled 200 points on each curve).
\tabname~\ref{tab:sym_eval} presents the quantitative metric results for the predicted paths with and without symmetry loss.
The model trained with symmetry loss outperformed the model trained without symmetry loss in maintaining the symmetry of the predicted paths.
\end{description}

\begin{table}
\begin{center}
\begin{tabular}{ c| c c } 
 \Xhline{1.2pt}
  & w/o symmetry loss &  w/ symmetry loss \\ \Xhline{1.2pt}
 SP symmetry & 3.46 & 1.21 \\ 
 CP symmetry & 1.08 & 0.34 \\ 
 \Xhline{1.2pt}
\end{tabular}
\caption{
Quantitative results for the predicted paths with and without symmetry loss.
In this table, we show the average value of the sample point (SP) and the control point (CP) symmetry error for the test set.
For both symmetry losses, the paths synthesized by network training with symmetry loss are lower.}
\label{tab:sym_eval}
\end{center}
\end{table}

\subsubsection{Effects of rendering accuracy loss}
\majorhl{
The rendering accuracy loss mainly aids in optimizing the correct color of the predicted path.
To evaluate the effect of the rendering accuracy loss effectively, we compared our complete synthesis method (with rendering accuracy loss) with an alternative version of it called \textit{NoRender}, in which we directly predicted the RGB values by using a linear layer with three outputs.
Then, we generated 10000 random paths with filled color and rendered them into $64\times64$ raster images. 
Of the 10000 generated images, 9000 formed the training set and 1000 formed the test set.
\mbox{\figname~\ref{fig:ablation}}(c) indicates that the path synthesized without the rendering accuracy loss could not achieve the correct color, and the curve positions of this path were not as accurate as those of the synthesized path with rendering accuracy loss.
}
\subsubsection{Effects of shape loss combination}
\majorhl{
We used two types of shape losses (\ie~the ordered Chamfer loss and earth Mover's loss) in our loss function.
We followed the same process of evaluating the effect of each loss by generating 10000 random paths without filled colors (9000 images formed the training set and 1000 images formed the test set).
We compared three versions of shape loss: ordered Chamfer loss only, earth Mover's loss only, and combined Chamfer and earth mover's loss. 
\mbox{\tabname~\ref{tab:shape_loss}} presents the relevant quantitative results.
The combined loss function achieved the smallest geometric differences; however, the earth Mover's loss only provided a marginal improvement.
}
\begin{table}
\begin{center}
\begin{tabular}{ c| c c c} 
 \Xhline{1.2pt}
   & ordered Chamfer & Earth Mover's & Combined \\ \Xhline{1.2pt}
 average distances & 1.85 & 3.67 & 1.12 \\ 
 \Xhline{1.2pt}
\end{tabular}
\caption{
Average control point distances between the input path and the predicted paths for three types of shape losses (the combined shape loss achieved the smallest shape distances).
}
\label{tab:shape_loss}
\end{center}
\end{table}



\subsubsection{Effects of curriculum training}
\begin{figure}
\includegraphics[width=\linewidth]{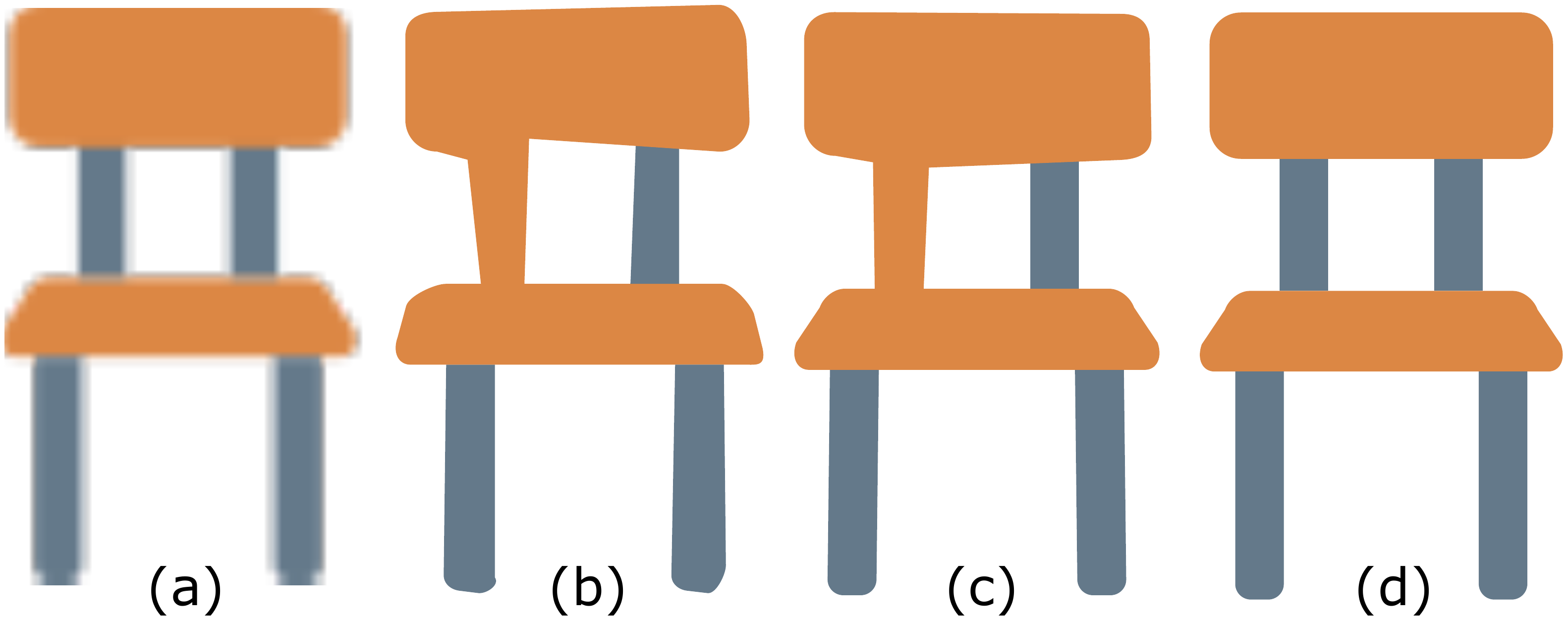}
\caption{
Vectorization results of different training strategies.
Given (a) the input raster clipart, the result of TS1 (b) fails to reconstruct some of the shapes and combine two separate paths due to the lack of the layering knowledge introduced in the second stage of training. With the introduction of layering knowledge, the result of TS2 (c) improves and better aids in reconstructing each closed path. 
(d) The full curriculum training strategy (TS3) achieved the best vectorization result.
}
\label{fig:ablation_training}
\end{figure}

Our curriculum training strategy can gradually improve the vectorization quality.
We evaluated our curriculum training strategy's effects numerically by computing the average image difference between the raster input clipart and the rasterized result of the vectorized clipart by using the proposed method (\ie~we wanted the vector clipart to represent the input raster clipart accurately).
In particular, we normalized the raster image to $0-1$ and used the total image difference between the two images divided by the total pixel number.
We used 20 clipart images in the test set of \textit{ClipNet} and rasterized them as the input.
We compared the following three training strategies (TSs):
\begin{itemize}
    \item TS1: Only the final stage was used.
    \item TS2: Only the second and final stages were used.
    \item TS3: all three stages were used.
\end{itemize}
\tabname~\ref{tab:training} presents the quantitative metric results, and \figname~\ref{fig:ablation_training} illustrates the vectorization results obtained by different training strategies.
Our curriculum training scheme achieved the smallest image difference and thus obtained the best vectorization quality.

\begin{table}
\begin{center}
\begin{tabular}{ c| c c c} 
 \Xhline{1.2pt}
   & TS1 &  TS2 & TS3 \\ \Xhline{1.2pt}
 average image difference & 0.374 & 0.266 & 0.112 \\ 
 \Xhline{1.2pt}
\end{tabular}
\caption{
Average image difference between the input raster clipart and the rasterized vector clipart synthesized by our method (the full curriculum training strategy (TS3) achieved the smallest difference).
}
\label{tab:training}
\end{center}
\end{table}

\subsection{Application}
\begin{figure*}
\includegraphics[width=\linewidth]{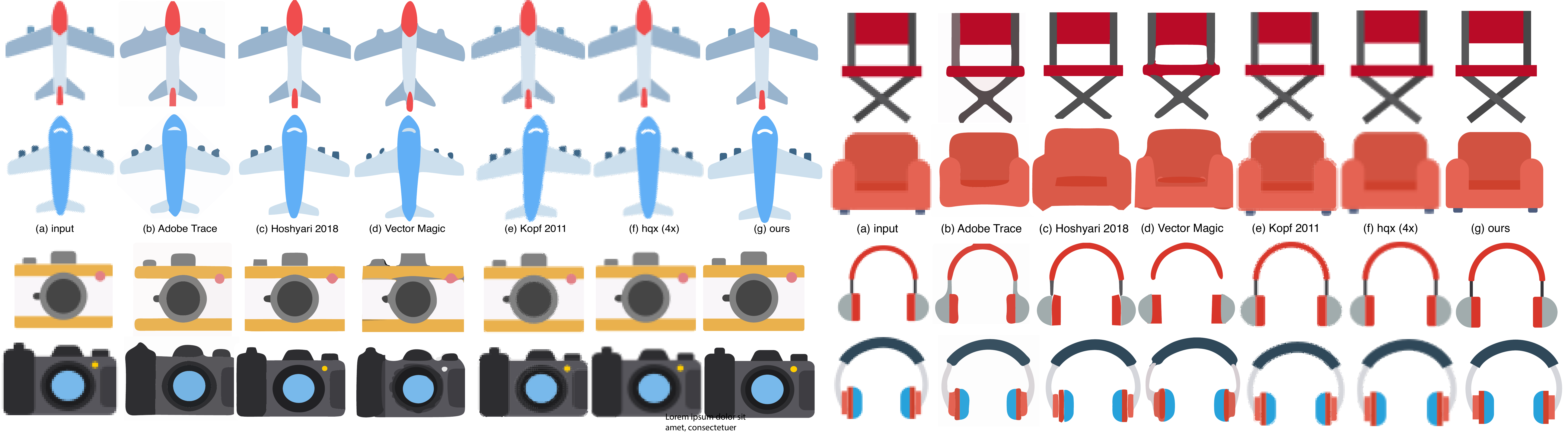}
\caption{
Vectorization of airplane, camera, chair, and headphones clipart.
Given the input clipart (a), we compared our result (e) with those obtained using (b) Adobe Trace~\cite{illustrator}, (c) Hoshyari~\etal~\cite{hoshyari2018perception}, (d) Vector Magic~\cite{vector_magic}, (e) a depixelization method~\cite{kopf2011depixelizing}, and (f) a popular pixel-art upscaling method~\cite{hqx}. 
}
\label{fig:four_vec_result}
\end{figure*}

\begin{figure}
\includegraphics[width=\linewidth]{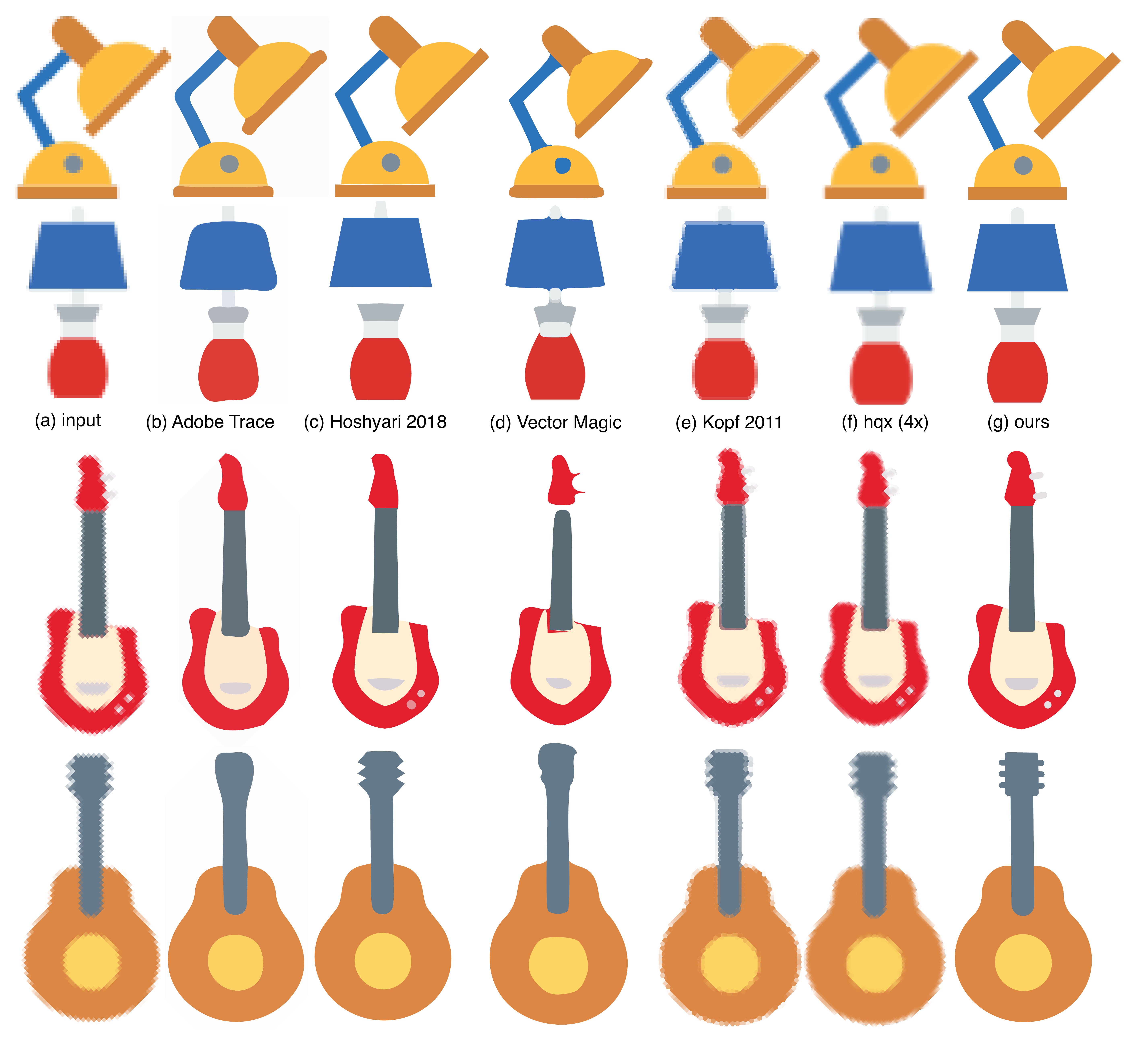}
\caption{
Vectorization of lamp and guitar clipart.
Given the input clipat (a), we compared our result (e) with those obtained using (b) Adobe Trace~\cite{illustrator}, (c) Hoshyari~\etal~\cite{hoshyari2018perception}, (d) Vector Magic~\cite{vector_magic}, (e) a depixelization method~\cite{kopf2011depixelizing}, and (f) a popular pixel-art upscaling method~\cite{hqx}. 
}
\label{fig:two_vec_result}
\end{figure}

\begin{figure}[h!]
\includegraphics[width=\linewidth]{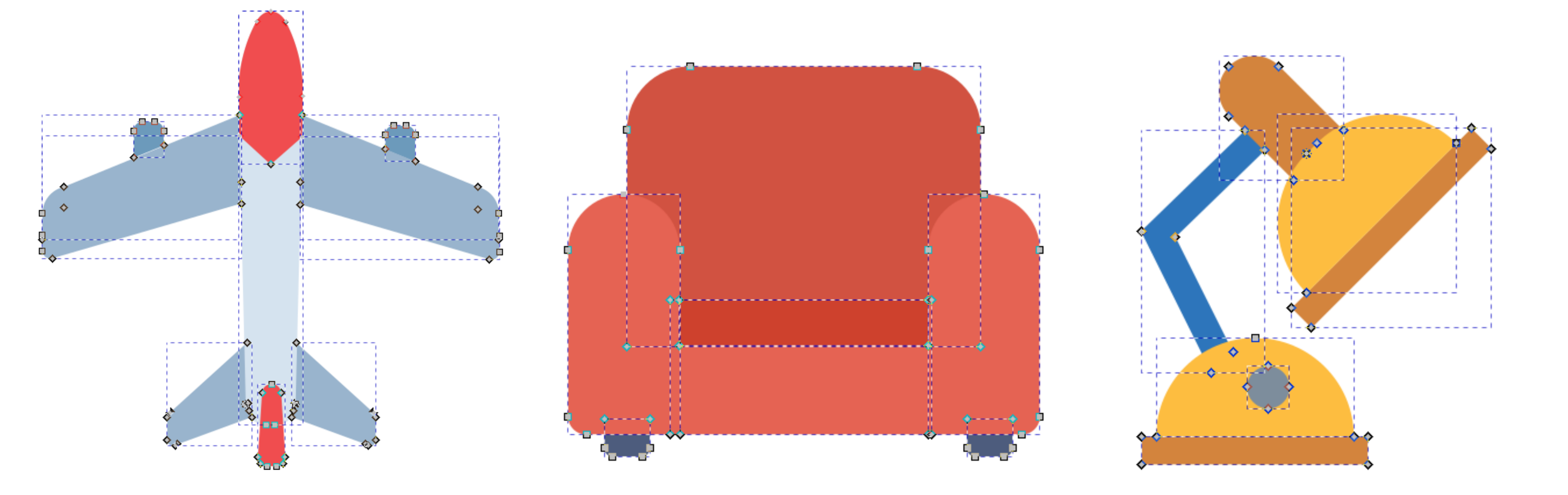}
\caption{We show the control points of several vectorization results to demonstrate the editability of our method.
In each shape, the gray points are the control points.
The control points generated by our method exhibit symmetry property to enhance editability.
}
\label{fig:edit_viz}
\end{figure}

\begin{figure}
\centering
\includegraphics[width=\linewidth]{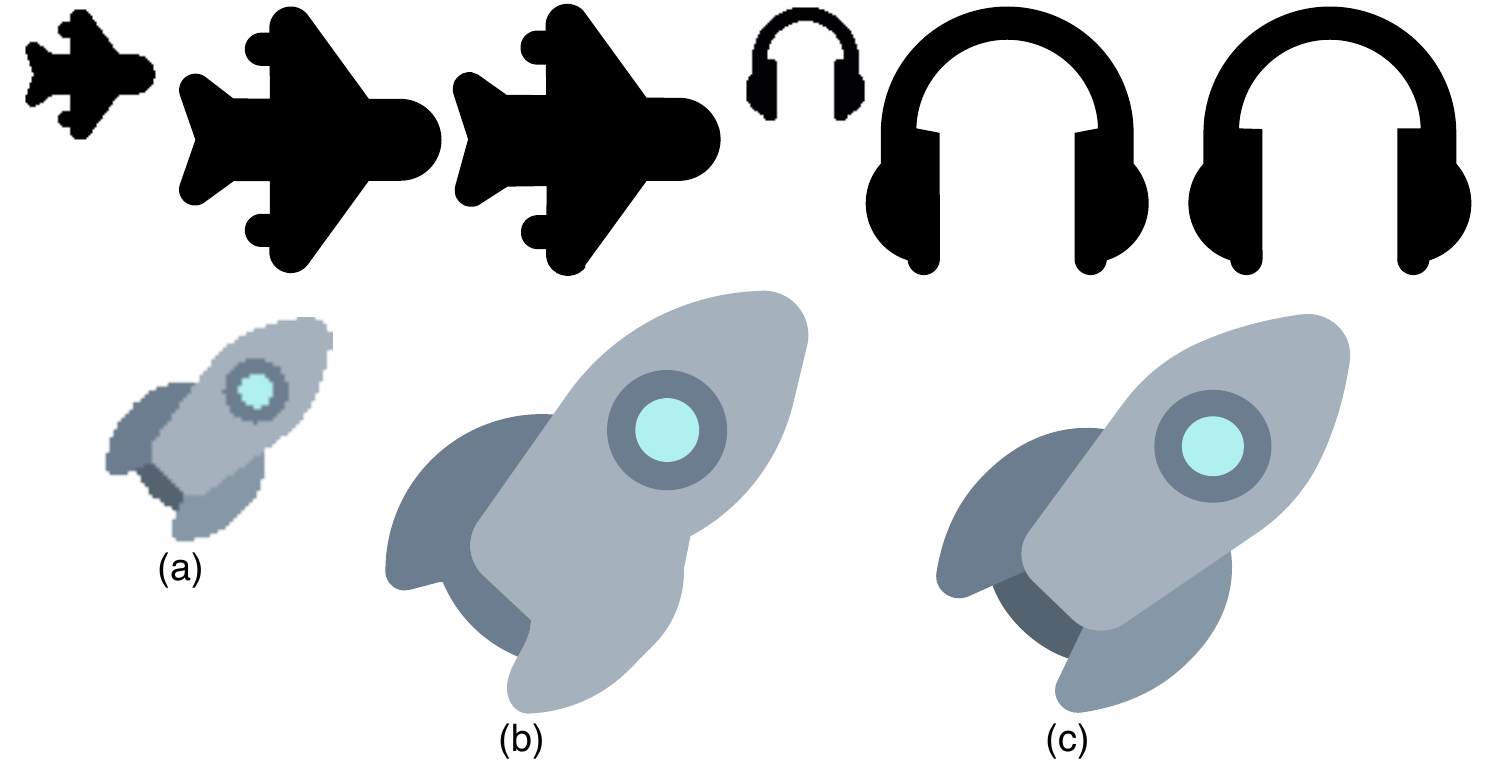}
\caption{
Given the input aliased clipart, we compared (b) the result of  Hoshyari~\etal~\cite{hoshyari2018perception} with (c) our result.
}
\label{fig:compare_percep}
\end{figure}

\begin{figure*}
\includegraphics[width=\linewidth]{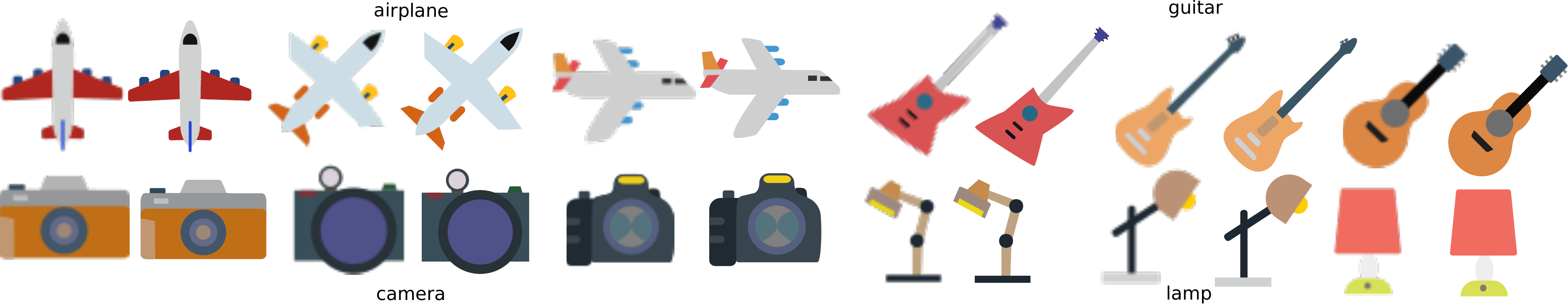}
\caption{
\majorhl{
Synthesized cliparts for airplanes, guitars, cameras, and lamps.
In each example, the left image indicates the synthesized raster clipart (as target shape) and the right image indicates the vector clipart synthesized by our method.
}
}
\label{fig:first_syn}
\end{figure*}

\begin{figure*}
\includegraphics[width=\linewidth]{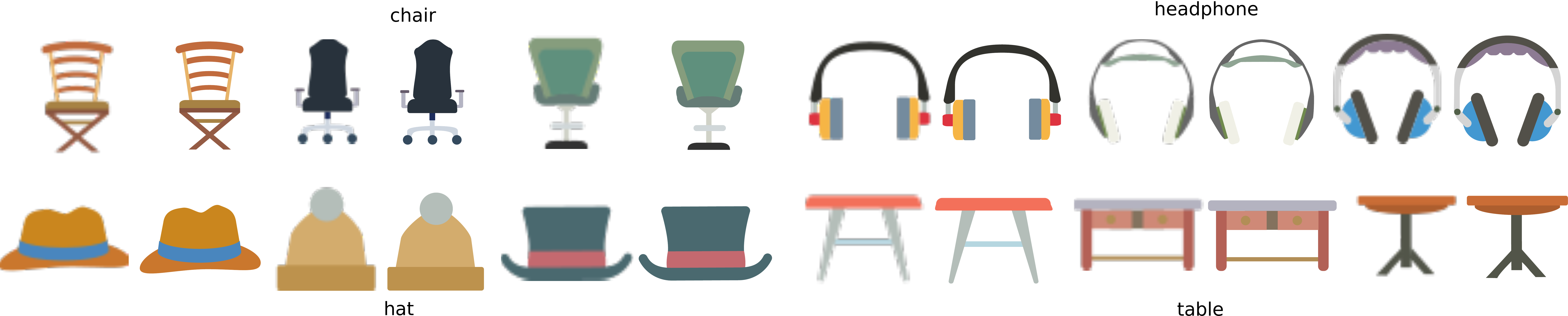}
\caption{
\majorhl{
Synthesized cliparts for chairs, headphones, hats, and tables.
In each example, the left image indicates the synthesized raster clipart (as target shape) and the right image indicates the vector clipart synthesized by our method.
}
}
\label{fig:second_syn}
\end{figure*}

\begin{figure}
\includegraphics[width=\linewidth]{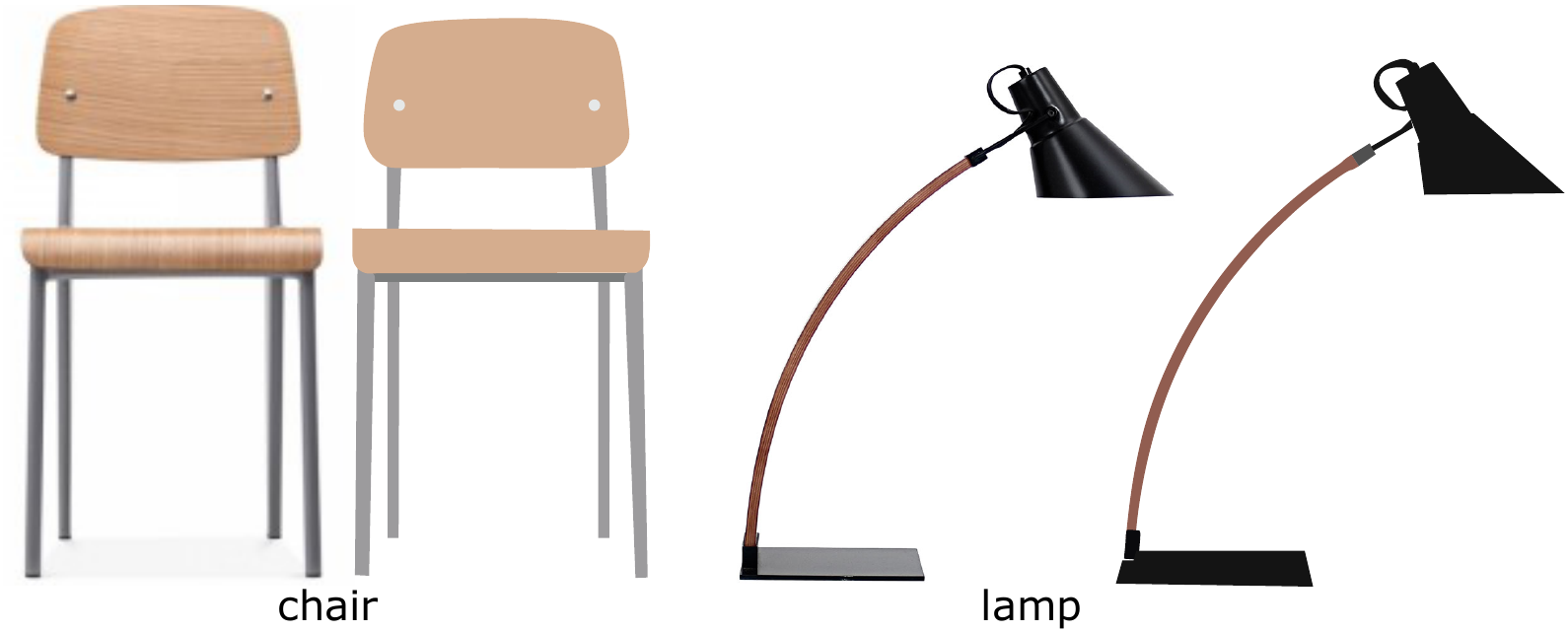}
\caption{
Examples of clipart generated from photographs.
For each example, the left image indicates the input raster photograph and the right image indicates the synthesized clipart generated using our approach.
}
\label{fig:photo_result}
\end{figure}

Our generative model supports various applications, including raster clipart vectorization, vector clipart synthesis, and photo-to-clipart conversion.
\begin{description}[style=unboxed,leftmargin=0cm]
\item[Raster clipart vectorization]
The most straightforward application of our generative model is to vectorize the artist-drawn clipart stored in the raster format.
Our goal was similar to that in \cite{hoshyari2018perception}; however, our method can process an anti-aliased input, and we first provide comparison against to this method using anti-aliased input.
We compared our results with those obtained with a range of available methods, including two publicly available vectorization methods, namely Adobe Trace~\cite{illustrator} and Vector Magic~\cite{vector_magic}.
Moreover, we tested the proposed method against state-of-the-art upscaling~\cite{hqx} and depixelization methods~\cite{kopf2011depixelizing}.
All the results are presented in \figname~\ref{fig:four_vec_result} and \figname~\ref{fig:two_vec_result}.
\figname~\ref{fig:compare_percep} presents a comparison of our method with a perceptual-based method~\cite{hoshyari2018perception} using aliased input.
We converted the input vector clipart into an aliased raster image by using Inkscape~\cite{Inkscape} and performed vectorization on the aliased clipart.
Our method generated a regular vector clipart despite the aliasing conversion, whereas the perceptual-based method \cite{hoshyari2018perception} contained some irregular details due to the rasterized clipart.
In \figname~\ref{fig:edit_viz}, we also show the control points of several synthesized results to demonstrate the editability of our method.

\item[Vector clipart synthesis]
For clipart synthesis, we first synthesized various images containing different categories of clipart. 
Then, we synthesized novel clipart by using our generative model by adopting the aforementioned images as target images.
Specifically, we used the training data in \textit{ClipNet} to train a conditional-GAN~\cite{condGAN} with the Wasserstein-GAN~\cite{wgan} (WGAN) training strategy with gradient penalty~\cite{imp_wgan}.
After completing the training, we can synthesized new clipart images of different categories.
The results obtained for different categories are presented in \figname~\ref{fig:first_syn} and \figname~\ref{fig:second_syn}, where the synthesized raster target shape for each clipart is presented on the left and the vector clipart is presented on the right.

\item[Photo-to-clipart]
Another application of our method is to synthesize the clipart of an input image $I$.
The goal is not to reproduce accurately the input image $I$ pixel-by-pixel but to generate perceived clipart similar to the input image.
For this purpose, we modified the loss function of the rendering accuracy loss from the raw image difference to the perceptual loss~\cite{johnson2016perceptual}.
In particular, we used perceptual loss on VGG16 features, which is expressed as follows:
\begin{align}
L_{perceptual}(I, \hat{I}) = \sum_{j=1}^{4}\frac{1}{N_j}\|\phi_j(I) - \phi_j(\hat{I})\|_2^2,
\end{align}
where $\hat{I}$ is the rendered image of the predicted clipart, and $\phi_j$ denotes the feature output of the VGG16~\cite{simonyan2014very} layers.
We used four layers, including \textit{conv1\_1}, \textit{conv1\_2}, \textit{conv3\_2}, and \textit{conv4\_2}.
\majorhl{
We prepared the training data for this application by improving the realism of clipart in \textit{ClipNet} using \mbox{\cite{shrivastava2017learning}}.
In short, we prepared a real photograph dataset for each category and used the adversarial training to make the clipart more photo-realistic.
In addition, we used the same training process as that followed for other applications.
}
The converted clipart obtained from photos is presented in \figname~\ref{fig:photo_result}.
We can observe that our method can synthesize clipart that captures the structure of the object in the input photograph.
\end{description}

\section{Conclusions}
This paper proposes an iterative synthesis model for synthesizing vector clipart that can be used for raster clipart vectorization, vector clipart synthesis, and photo-to-clipart conversion. 
The key technical contribution of this paper is that we formulated the clipart synthesis problem as an iterative two-step process by using two separate modules; namely the \textit{``Continue to add a layer?''} and \textit{``What path to add next?''} modules.
Then, we used a recurrent decoder to synthesize the curves sequentially by using joint loss functions (shape loss, symmetry loss, smoothness loss, and rendering accuracy
loss). 
In addition, we developed a dataset called \textit{ClipNet}, which is a collection of clipart of man-made objects, to facilitate the training of the proposed synthesis model.
This dataset currently contains clipart on ten categories of man-made objects and would continue to grow as additional clipart is uploaded to its web repositories. 
We evaluated the behavior and quality of the synthesized clipart on various aspects. 
We demonstrated that our synthesis model can vectorize and synthesize man-made object' clipart that can be recognized by humans. 
Our synthesis model can synthesize clipart belonging to the ten categories of the ClipNet dataset, which is common for the previous synthesis model~\cite{kalogerakis2012probabilistic}.
Finally, we will make the aforementioned dataset public in the future, and we believe that it would benefit future research on data-driven vector graphics.


\subsection{Limitations}
Due to the complexity of the two-modules synthesis model (especially the second module), our synthesis model can not predict a new layer in real-time. 
Thus, simpler architectures (\eg~those with a simpler recurrent decoder) to facilitate faster inference are worth investigating.
In addition, we only used the constant color filling model in our vector rendering model. 
Although this model is sufficient for modeling considerable clipart, some clipart on the Internet requires more complex color models, such as linear or radial gradient models, or a complicated shading method~\cite{lopez2013depicting}. 
We plan to investigate how to support the linear gradient in our recurrent decoder to generate detailed clipart.
A differentiable vector graphics rendering technique supports additional primitives, stroke styles, and color models, which would considerably benefit our synthesis model.

Moreover, our synthesis model does not consider the different viewpoints of clipart (\eg~the top, side, and front views of an airplane) during training and inference.
Numerous types of clipart are created under different viewpoints. 
Our synthesis model sometimes yields inappropriate clipart results that combine structures from different viewpoints.
In the future, we will attempt to enrich our dataset with viewpoint information and design a novel architecture that considers the viewpoint information.
Finally, we will extend our dataset with different clipart styles (as explored in \cite{GarcesSIG2014}) and design a framework to vectorize and synthesize clipart in various styles.




\ifCLASSOPTIONcompsoc
  \section*{Acknowledgments}
\else
  \section*{Acknowledgment}
\fi

This work was supported in part by the Ministry of Science and Technology, Taiwan, under Grant MOST109-2218-E-002-030, 109-2634-F-002-032, and National Taiwan University.
And we are grateful to the National Center for High-performance Computing.
We want to thank Tzu-mao Li, Sheng-Jie Luo, Yu-Ting Wu, Chi-Lan Yang, and anonymous reviewers for insightful suggestions and discussion.
I-Chao Shen was supported by the MediaTek Fellowship.

\ifCLASSOPTIONcaptionsoff
  \newpage
\fi


\bibliographystyle{IEEEtran}
\bibliography{paper.bib}

\begin{thebibliography}{10}
\providecommand{\url}[1]{#1}
\csname url@samestyle\endcsname
\providecommand{\newblock}{\relax}
\providecommand{\bibinfo}[2]{#2}
\providecommand{\BIBentrySTDinterwordspacing}{\spaceskip=0pt\relax}
\providecommand{\BIBentryALTinterwordstretchfactor}{4}
\providecommand{\BIBentryALTinterwordspacing}{\spaceskip=\fontdimen2\font plus
\BIBentryALTinterwordstretchfactor\fontdimen3\font minus
  \fontdimen4\font\relax}
\providecommand{\BIBforeignlanguage}[2]{{%
\expandafter\ifx\csname l@#1\endcsname\relax
\typeout{** WARNING: IEEEtran.bst: No hyphenation pattern has been}%
\typeout{** loaded for the language `#1'. Using the pattern for}%
\typeout{** the default language instead.}%
\else
\language=\csname l@#1\endcsname
\fi
#2}}
\providecommand{\BIBdecl}{\relax}
\BIBdecl

\bibitem{illustrator}
\BIBentryALTinterwordspacing
Adobe, ``Adobe illustrator 2020: Image trace,'' 2020. [Online]. Available:
  \url{http://www.adobe.com/}
\BIBentrySTDinterwordspacing

\bibitem{Inkscape}
\BIBentryALTinterwordspacing
{Inkscape Project}, ``Inkscape,'' 2020. [Online]. Available:
  \url{https://inkscape.org}
\BIBentrySTDinterwordspacing

\bibitem{karras2019analyzing}
T.~Karras, S.~Laine, M.~Aittala, J.~Hellsten, J.~Lehtinen, and T.~Aila,
  ``Analyzing and improving the image quality of stylegan,'' \emph{arXiv
  preprint arXiv:1912.04958}, 2019.

\bibitem{kalogerakis2012probabilistic}
E.~Kalogerakis, S.~Chaudhuri, D.~Koller, and V.~Koltun, ``A probabilistic model
  for component-based shape synthesis,'' \emph{ACM Transactions on Graphics
  (TOG)}, vol.~31, no.~4, p.~55, 2012.

\bibitem{wang2018deep}
K.~Wang, M.~Savva, A.~X. Chang, and D.~Ritchie, ``Deep convolutional priors for
  indoor scene synthesis,'' \emph{ACM Transactions on Graphics (TOG)}, vol.~37,
  no.~4, p.~70, 2018.

\bibitem{Li:2020:DVG}
T.-M. Li, M.~Luk\'{a}\v{c}, G.~Micha\"{e}l, and J.~Ragan-Kelley,
  ``Differentiable vector graphics rasterization for editing and learning,''
  \emph{ACM Trans. Graph. (Proc. SIGGRAPH Asia)}, vol.~39, no.~6, pp.
  193:1--193:15, 2020.

\bibitem{imagenet_cvpr09}
J.~Deng, W.~Dong, R.~Socher, L.-J. Li, K.~Li, and L.~Fei-Fei, ``{ImageNet: A
  Large-Scale Hierarchical Image Database},'' in \emph{CVPR09}, 2009.

\bibitem{chang2015shapenet}
A.~X. Chang, T.~Funkhouser, L.~Guibas, P.~Hanrahan, Q.~Huang, Z.~Li,
  S.~Savarese, M.~Savva, S.~Song, H.~Su \emph{et~al.}, ``Shapenet: An
  information-rich 3d model repository,'' \emph{arXiv preprint
  arXiv:1512.03012}, 2015.

\bibitem{vector_magic}
\BIBentryALTinterwordspacing
{Vector Magic}, ``Cedar lake ventures,'' 2020. [Online]. Available:
  \url{https://vectormagic.com/}
\BIBentrySTDinterwordspacing

\bibitem{FLB17}
\BIBentryALTinterwordspacing
J.-D. Favreau, F.~Lafarge, and A.~Bousseau, ``Photo2clipart: Image abstraction
  and vectorization using layered linear gradients,'' \emph{ACM Transactions on
  Graphics (SIGGRAPH Asia Conference Proceedings)}, vol.~36, no.~6, November
  2017. [Online]. Available:
  \url{http://www-sop.inria.fr/reves/Basilic/2017/FLB17}
\BIBentrySTDinterwordspacing

\bibitem{orzan2008diffusion}
A.~Orzan, A.~Bousseau, H.~Winnem{\"o}ller, P.~Barla, J.~Thollot, and
  D.~Salesin, ``Diffusion curves: a vector representation for smooth-shaded
  images,'' in \emph{ACM Transactions on Graphics (TOG)}, vol.~27, no.~3.\hskip
  1em plus 0.5em minus 0.4em\relax ACM, 2008, p.~92.

\bibitem{sun2007image}
J.~Sun, L.~Liang, F.~Wen, and H.-Y. Shum, ``Image vectorization using optimized
  gradient meshes,'' in \emph{ACM Transactions on Graphics (TOG)}, vol.~26,
  no.~3.\hskip 1em plus 0.5em minus 0.4em\relax ACM, 2007, p.~11.

\bibitem{xia2009patch}
T.~Xia, B.~Liao, and Y.~Yu, ``Patch-based image vectorization with automatic
  curvilinear feature alignment,'' \emph{ACM Transactions on Graphics (TOG)},
  vol.~28, no.~5, p. 115, 2009.

\bibitem{kim2018semantic}
B.~Kim, O.~Wang, A.~C. {\"O}ztireli, and M.~Gross, ``Semantic segmentation for
  line drawing vectorization using neural networks,'' in \emph{Computer
  Graphics Forum}, vol.~37, no.~2.\hskip 1em plus 0.5em minus 0.4em\relax Wiley
  Online Library, 2018, pp. 329--338.

\bibitem{bessmeltsev2019vectorization}
M.~Bessmeltsev and J.~Solomon, ``Vectorization of line drawings via polyvector
  fields,'' \emph{ACM Transactions on Graphics (TOG)}, vol.~38, no.~1, p.~9,
  2019.

\bibitem{Liu:2016:DI}
Y.~Liu, A.~Agarwala, J.~Lu, and S.~Rusinkiewicz, ``Data-driven iconification,''
  in \emph{International Symposium on Non-Photorealistic Animation and
  Rendering (NPAR)}, May 2016.

\bibitem{xie2017}
X.~Jun, W.~Holger, L.~Wilmot, and S.~Stephen, ``Interactive vectorization,'' in
  \emph{Proceedings of SIGCHI 2017}.\hskip 1em plus 0.5em minus 0.4em\relax
  ACM, 2017.

\bibitem{hqx}
\BIBentryALTinterwordspacing
M.~Stepin, ``Hqx,'' 2003. [Online]. Available:
  \url{http://web.archive.org/web/20070717064839/www.hiend3d.com/hq4x.html}
\BIBentrySTDinterwordspacing

\bibitem{kopf2011depixelizing}
J.~Kopf and D.~Lischinski, ``Depixelizing pixel art,'' \emph{ACM Transactions
  on graphics (TOG)}, vol.~30, no.~4, p.~99, 2011.

\bibitem{hoshyari2018perception}
S.~Hoshyari, E.~A. Dominici, A.~Sheffer, N.~Carr, Z.~Wang, D.~Ceylan, I.~Shen
  \emph{et~al.}, ``Perception-driven semi-structured boundary vectorization,''
  \emph{ACM Transactions on Graphics (TOG)}, vol.~37, no.~4, p. 118, 2018.

\bibitem{chaudhuri2011probabilistic}
S.~Chaudhuri, E.~Kalogerakis, L.~Guibas, and V.~Koltun, ``Probabilistic
  reasoning for assembly-based 3d modeling,'' in \emph{ACM Transactions on
  Graphics (TOG)}, vol.~30, no.~4.\hskip 1em plus 0.5em minus 0.4em\relax ACM,
  2011, p.~35.

\bibitem{fisher2012example}
M.~Fisher, D.~Ritchie, M.~Savva, T.~Funkhouser, and P.~Hanrahan,
  ``Example-based synthesis of 3d object arrangements,'' \emph{ACM Transactions
  on Graphics (TOG)}, vol.~31, no.~6, p. 135, 2012.

\bibitem{lin2013probabilistic}
S.~Lin, D.~Ritchie, M.~Fisher, and P.~Hanrahan, ``Probabilistic
  color-by-numbers: Suggesting pattern colorizations using factor graphs,''
  \emph{ACM Transactions on Graphics (TOG)}, vol.~32, no.~4, p.~37, 2013.

\bibitem{kingma2013auto}
D.~P. Kingma and M.~Welling, ``Auto-encoding variational bayes,'' \emph{arXiv
  preprint arXiv:1312.6114}, 2013.

\bibitem{goodfellow2014generative}
I.~Goodfellow, J.~Pouget-Abadie, M.~Mirza, B.~Xu, D.~Warde-Farley, S.~Ozair,
  A.~Courville, and Y.~Bengio, ``Generative adversarial nets,'' in
  \emph{Advances in neural information processing systems}, 2014, pp.
  2672--2680.

\bibitem{pix2pix2016}
P.~Isola, J.-Y. Zhu, T.~Zhou, and A.~A. Efros, ``Image-to-image translation
  with conditional adversarial networks,'' \emph{arxiv}, 2016.

\bibitem{CycleGAN2017}
J.-Y. Zhu, T.~Park, P.~Isola, and A.~A. Efros, ``Unpaired image-to-image
  translation using cycle-consistent adversarial networks,'' in \emph{Computer
  Vision (ICCV), 2017 IEEE International Conference on}, 2017.

\bibitem{li2017grass}
J.~Li, K.~Xu, S.~Chaudhuri, E.~Yumer, H.~Zhang, and L.~Guibas, ``Grass:
  Generative recursive autoencoders for shape structures,'' \emph{ACM
  Transactions on Graphics (TOG)}, vol.~36, no.~4, p.~52, 2017.

\bibitem{zhu2019scores}
C.~Zhu, K.~Xu, S.~Chaudhuri, R.~Yi, and H.~Zhang, ``Scores: Shape composition
  with recursive substructure priors,'' \emph{ACM Transactions on Graphics
  (TOG)}, vol.~37, no.~6, p. 211, 2019.

\bibitem{li2019grains}
M.~Li, A.~G. Patil, K.~Xu, S.~Chaudhuri, O.~Khan, A.~Shamir, C.~Tu, B.~Chen,
  D.~Cohen-Or, and H.~Zhang, ``Grains: Generative recursive autoencoders for
  indoor scenes,'' \emph{ACM Transactions on Graphics (TOG)}, vol.~38, no.~2,
  p.~12, 2019.

\bibitem{zou20173d}
C.~Zou, E.~Yumer, J.~Yang, D.~Ceylan, and D.~Hoiem, ``3d-prnn: Generating shape
  primitives with recurrent neural networks,'' in \emph{The IEEE International
  Conference on Computer Vision (ICCV)}, 2017.

\bibitem{Yan:2019:RRP}
\BIBentryALTinterwordspacing
Z.~Yan, R.~Hu, X.~Yan, L.~Chen, O.~Van~Kaick, H.~Zhang, and H.~Huang,
  ``Rpm-net: Recurrent prediction of motion and parts from point cloud,''
  \emph{ACM Trans. Graph.}, vol.~38, no.~6, pp. 240:1--240:15, Nov. 2019.
  [Online]. Available: \url{http://doi.acm.org/10.1145/3355089.3356573}
\BIBentrySTDinterwordspacing

\bibitem{ha2017neural}
D.~Ha and D.~Eck, ``A neural representation of sketch drawings,'' \emph{arXiv
  preprint arXiv:1704.03477}, 2017.

\bibitem{zheng2019content}
X.~Zheng, X.~Qiao, Y.~Cao, and R.~W. Lau, ``Content-aware generative modeling
  of graphic design layouts,'' \emph{ACM Transactions on Graphics (TOG)},
  vol.~38, no.~4, p. 133, 2019.

\bibitem{li2018layoutgan}
\BIBentryALTinterwordspacing
J.~Li, T.~Xu, J.~Zhang, A.~Hertzmann, and J.~Yang, ``Layout{GAN}: Generating
  graphic layouts with wireframe discriminator,'' in \emph{International
  Conference on Learning Representations}, 2019. [Online]. Available:
  \url{https://openreview.net/forum?id=HJxB5sRcFQ}
\BIBentrySTDinterwordspacing

\bibitem{acuna2018efficient}
D.~Acuna, H.~Ling, A.~Kar, and S.~Fidler, ``Efficient interactive annotation of
  segmentation datasets with polygon-rnn++,'' in \emph{Proceedings of the IEEE
  Conference on Computer Vision and Pattern Recognition}, 2018, pp. 859--868.

\bibitem{han-2018-pixelization}
C.~Han, Q.~Wen, S.~He, Q.~Zhu, Y.~Tan, G.~Han, and T.-T. Wong, ``Deep
  unsupervised pixelization,'' \emph{ACM Transactions on Graphics (SIGGRAPH
  Asia 2018 issue)}, vol.~37, no.~6, pp. 243:1--243:11, November 2018.

\bibitem{Lopes_2019_ICCV}
R.~G. Lopes, D.~Ha, D.~Eck, and J.~Shlens, ``A learned representation for
  scalable vector graphics,'' in \emph{The IEEE International Conference on
  Computer Vision (ICCV)}, October 2019.

\bibitem{sbai2018vector}
O.~Sbai, C.~Couprie, and M.~Aubry, ``Vector image generation by learning
  parametric layer decomposition,'' \emph{arXiv preprint arXiv:1812.05484},
  2018.

\bibitem{garmentdesign_Wang_SA18}
T.~Y. Wang, D.~Ceylan, J.~Popovic, and N.~J. Mitra, ``Learning a shared shape
  space for multimodal garment design,'' \emph{ACM Trans. Graph.}, vol.~37,
  no.~6, pp. 1:1--1:14, 2018.

\bibitem{guerin2017interactive}
E.~Gu{\'e}rin, J.~Digne, E.~Galin, A.~Peytavie, C.~Wolf, B.~Benes, and
  B.~Martinez, ``Interactive example-based terrain authoring with conditional
  generative adversarial networks,'' \emph{Acm Transactions on Graphics (TOG)},
  vol.~36, no.~6, p. 228, 2017.

\bibitem{Bau:Ganpaint:2019}
D.~Bau, H.~Strobelt, W.~Peebles, J.~Wulff, B.~Zhou, J.~Zhu, and A.~Torralba,
  ``Semantic photo manipulation with a generative image prior,'' \emph{ACM
  Transactions on Graphics (Proceedings of ACM SIGGRAPH)}, vol.~38, no.~4,
  2019.

\bibitem{hin2019interactive}
T.~C.~L. Hin, I.~Shen, I.~Sato, T.~Igarashi \emph{et~al.}, ``Interactive
  subspace exploration on generative image modelling,'' \emph{arXiv preprint
  arXiv:1906.09840}, 2019.

\bibitem{zhu2016generative}
J.-Y. Zhu, P.~Kr{\"a}henb{\"u}hl, E.~Shechtman, and A.~A. Efros, ``Generative
  visual manipulation on the natural image manifold,'' in \emph{Proceedings of
  European Conference on Computer Vision (ECCV)}, 2016.

\bibitem{lin2014microsoft}
T.-Y. Lin, M.~Maire, S.~Belongie, J.~Hays, P.~Perona, D.~Ramanan,
  P.~Doll{\'a}r, and C.~L. Zitnick, ``Microsoft coco: Common objects in
  context,'' in \emph{European conference on computer vision}.\hskip 1em plus
  0.5em minus 0.4em\relax Springer, 2014, pp. 740--755.

\bibitem{liu2015deep}
Z.~Liu, P.~Luo, X.~Wang, and X.~Tang, ``Deep learning face attributes in the
  wild,'' in \emph{Proceedings of the IEEE international conference on computer
  vision}, 2015, pp. 3730--3738.

\bibitem{Perazzi2016}
F.~Perazzi, J.~Pont-Tuset, B.~McWilliams, L.~{Van Gool}, M.~Gross, and
  A.~Sorkine-Hornung, ``A benchmark dataset and evaluation methodology for
  video object segmentation,'' in \emph{Computer Vision and Pattern
  Recognition}, 2016.

\bibitem{Wu_2015_CVPR}
Z.~Wu, S.~Song, A.~Khosla, F.~Yu, L.~Zhang, X.~Tang, and J.~Xiao, ``3d
  shapenets: A deep representation for volumetric shapes,'' in
  \emph{Proceedings of the IEEE Conference on Computer Vision and Pattern
  Recognition (CVPR)}, June 2015.

\bibitem{Mo_2019_CVPR}
K.~Mo, S.~Zhu, A.~X. Chang, L.~Yi, S.~Tripathi, L.~J. Guibas, and H.~Su,
  ``{PartNet}: A large-scale benchmark for fine-grained and hierarchical
  part-level {3D} object understanding,'' in \emph{The IEEE Conference on
  Computer Vision and Pattern Recognition (CVPR)}, June 2019.

\bibitem{Koch_2019_CVPR}
S.~Koch, A.~Matveev, Z.~Jiang, F.~Williams, A.~Artemov, E.~Burnaev, M.~Alexa,
  D.~Zorin, and D.~Panozzo, ``Abc: A big cad model dataset for geometric deep
  learning,'' in \emph{The IEEE Conference on Computer Vision and Pattern
  Recognition (CVPR)}, June 2019.

\bibitem{song2016ssc}
S.~Song, F.~Yu, A.~Zeng, A.~X. Chang, M.~Savva, and T.~Funkhouser, ``Semantic
  scene completion from a single depth image,'' \emph{Proceedings of 30th IEEE
  Conference on Computer Vision and Pattern Recognition}, 2017.

\bibitem{shugrina2019creative}
M.~Shugrina, Z.~Liang, A.~Kar, J.~Li, A.~Singh, K.~Singh, and S.~Fidler,
  ``Creative flow+ dataset,'' in \emph{The IEEE Conference on Computer Vision
  and Pattern Recognition (CVPR)}, June 2019.

\bibitem{GSHPDB19}
\BIBentryALTinterwordspacing
Y.~Gryaditskaya, M.~Sypesteyn, J.~W. Hoftijzer, S.~Pont, F.~Durand, and
  A.~Bousseau, ``Opensketch: A richly-annotated dataset of product design
  sketches,'' \emph{ACM Transactions on Graphics (SIGGRAPH Asia Conference
  Proceedings)}, vol.~38, no.~6, November 2019. [Online]. Available:
  \url{http://www-sop.inria.fr/reves/Basilic/2019/GSHPDB19}
\BIBentrySTDinterwordspacing

\bibitem{eitz2012hdhso}
M.~Eitz, J.~Hays, and M.~Alexa, ``How do humans sketch objects?'' \emph{ACM
  Trans. Graph. (Proc. SIGGRAPH)}, vol.~31, no.~4, pp. 44:1--44:10, 2012.

\bibitem{Cole:2008:PDL}
\BIBentryALTinterwordspacing
F.~Cole, A.~Golovinskiy, A.~Limpaecher, H.~S. Barros, A.~Finkelstein,
  T.~Funkhouser, and S.~Rusinkiewicz, ``Where do people draw lines?'' \emph{ACM
  Trans. Graph.}, vol.~27, no.~3, pp. 88:1--88:11, Aug. 2008. [Online].
  Available: \url{http://doi.acm.org/10.1145/1360612.1360687}
\BIBentrySTDinterwordspacing

\bibitem{bell2013opensurfaces}
S.~Bell, P.~Upchurch, N.~Snavely, and K.~Bala, ``Opensurfaces: A richly
  annotated catalog of surface appearance,'' \emph{ACM Transactions on graphics
  (TOG)}, vol.~32, no.~4, p. 111, 2013.

\bibitem{long2015fully}
J.~Long, E.~Shelhamer, and T.~Darrell, ``Fully convolutional networks for
  semantic segmentation,'' in \emph{Proceedings of the IEEE conference on
  computer vision and pattern recognition}, 2015, pp. 3431--3440.

\bibitem{kingma2014adam}
D.~P. Kingma and J.~Ba, ``Adam: A method for stochastic optimization,'' 2014.

\bibitem{he2016deep}
K.~He, X.~Zhang, S.~Ren, and J.~Sun, ``Deep residual learning for image
  recognition,'' in \emph{Proceedings of the IEEE conference on computer vision
  and pattern recognition}, 2016, pp. 770--778.

\bibitem{hochreiter1997long}
S.~Hochreiter and J.~Schmidhuber, ``Long short-term memory,'' \emph{Neural
  computation}, vol.~9, no.~8, pp. 1735--1780, 1997.

\bibitem{fan2017point}
H.~Fan, H.~Su, and L.~J. Guibas, ``A point set generation network for 3d object
  reconstruction from a single image,'' in \emph{Proceedings of the IEEE
  conference on computer vision and pattern recognition}, 2017, pp. 605--613.

\bibitem{sorkine2004laplacian}
O.~Sorkine, D.~Cohen-Or, Y.~Lipman, M.~Alexa, C.~R{\"o}ssl, and H.-P. Seidel,
  ``Laplacian surface editing,'' in \emph{Proceedings of the 2004
  Eurographics/ACM SIGGRAPH symposium on Geometry processing}.\hskip 1em plus
  0.5em minus 0.4em\relax ACM, 2004, pp. 175--184.

\bibitem{bengio2009curriculum}
Y.~Bengio, J.~Louradour, R.~Collobert, and J.~Weston, ``Curriculum learning,''
  in \emph{Proceedings of the 26th annual international conference on machine
  learning}, 2009, pp. 41--48.

\bibitem{hacohen2019power}
G.~Hacohen and D.~Weinshall, ``On the power of curriculum learning in training
  deep networks,'' \emph{arXiv preprint arXiv:1904.03626}, 2019.

\bibitem{globfit}
\BIBentryALTinterwordspacing
Y.~Li, X.~Wu, Y.~Chrysathou, A.~Sharf, D.~Cohen-Or, and N.~J. Mitra, ``Globfit:
  Consistently fitting primitives by discovering global relations,'' vol.~30,
  no.~4, Jul. 2011. [Online]. Available:
  \url{https://doi.org/10.1145/2010324.1964947}
\BIBentrySTDinterwordspacing

\bibitem{paszke2017automatic}
A.~Paszke, S.~Gross, S.~Chintala, G.~Chanan, E.~Yang, Z.~DeVito, Z.~Lin,
  A.~Desmaison, L.~Antiga, and A.~Lerer, ``Automatic differentiation in
  {PyTorch},'' in \emph{NIPS Autodiff Workshop}, 2017.

\bibitem{He2016}
K.~{He}, X.~{Zhang}, S.~{Ren}, and J.~{Sun}, ``Deep residual learning for image
  recognition,'' in \emph{2016 IEEE Conference on Computer Vision and Pattern
  Recognition (CVPR)}, June 2016, pp. 770--778.

\bibitem{Ba2016LayerN}
J.~Ba, J.~R. Kiros, and G.~E. Hinton, ``Layer normalization,'' \emph{ArXiv},
  vol. abs/1607.06450, 2016.

\bibitem{Yin2017ComparativeSO}
W.~Yin, K.~Kann, M.~Yu, and H.~Sch{\"u}tze, ``Comparative study of cnn and rnn
  for natural language processing,'' \emph{ArXiv}, vol. abs/1702.01923, 2017.

\bibitem{condGAN}
\BIBentryALTinterwordspacing
M.~Mirza and S.~Osindero, ``Conditional generative adversarial nets,''
  \emph{CoRR}, vol. abs/1411.1784, 2014. [Online]. Available:
  \url{http://arxiv.org/abs/1411.1784}
\BIBentrySTDinterwordspacing

\bibitem{wgan}
M.~Arjovsky, S.~Chintala, and L.~Bottou, ``Wasserstein gan,'' \emph{arXiv
  preprint arXiv:1701.07875}, 2017.

\bibitem{imp_wgan}
I.~Gulrajani, F.~Ahmed, M.~Arjovsky, V.~Dumoulin, and A.~C. Courville,
  ``Improved training of wasserstein gans,'' in \emph{Advances in neural
  information processing systems}, 2017, pp. 5767--5777.

\bibitem{johnson2016perceptual}
J.~Johnson, A.~Alahi, and L.~Fei-Fei, ``Perceptual losses for real-time style
  transfer and super-resolution,'' in \emph{European conference on computer
  vision}.\hskip 1em plus 0.5em minus 0.4em\relax Springer, 2016, pp. 694--711.

\bibitem{simonyan2014very}
K.~Simonyan and A.~Zisserman, ``Very deep convolutional networks for
  large-scale image recognition,'' \emph{arXiv preprint arXiv:1409.1556}, 2014.

\bibitem{shrivastava2017learning}
A.~Shrivastava, T.~Pfister, O.~Tuzel, J.~Susskind, W.~Wang, and R.~Webb,
  ``Learning from simulated and unsupervised images through adversarial
  training,'' in \emph{Proceedings of the IEEE conference on computer vision
  and pattern recognition}, 2017, pp. 2107--2116.

\bibitem{lopez2013depicting}
J.~Lopez-Moreno, P.~Stefan, A.~Bousseau, M.~Agrawala, and G.~Drettakis,
  ``Depicting stylized materials with vector shade trees,'' 2013.

\bibitem{GarcesSIG2014}
E.~Garces, A.~Agarwala, D.~Gutierrez, and A.~Hertzmann, ``A similarity measure
  for illustration style,'' \emph{ACM Transactions on Graphics (SIGGRAPH
  2014)}, vol.~33, no.~4, 2014.

\end{thebibliography}

\end{document}


\maketitle
\input{network-structure}
\input{nearest_shape_viz}



